\documentclass[preprint1]{aastex}
\slugcomment{Submitted to {\it Astronomical Journal}, May 2001}

\begin{document}

\title{COLORS AND SPECTRA OF KUIPER BELT OBJECTS \altaffilmark{1}\\
}

\author{David C. Jewitt\altaffilmark{2}}
\affil{Institute for Astronomy\\
2680 Woodlawn Drive, Honolulu, HI 96822}

\and

\author{Jane X. Luu}
\affil{Sterrewacht Leiden\\
Postbus 9513,
2300RA Leiden, The Netherlands}

\altaffiltext{1}{Based [in part] on data collected at Subaru Telescope, 
which is operated by the National Astronomical Observatory of Japan.}

\altaffiltext{2}{Visiting Astronomer, W. M. Keck Observatory,
jointly operated by California Institute of Technology and the
University of California.}

\newpage

\begin{abstract}

We present new measurements of the optical colors of Kuiper Belt
Objects, principally from the Keck 10-m telescope.  The measurements
confirm the existence of a wide spread in the B-V, V-R and R-I color
indices (Luu and Jewitt 1996).  Relative to the sun, the Kuiper Belt
Objects exhibit reflected colors from nearly neutral to very red.  The
optical and optical-infrared (V-J) color indices are mutually
correlated, showing the presence of a single reddening agent from
$0.45\mu m$ to $1.2\mu m$.  On the other hand, we find no evidence for
linear correlations between the color and absolute magnitude (a proxy for
size), instantaneous heliocentric distance, semi-major axis, or with
any other orbital property.  In this regard, the Kuiper Belt Objects
differ from the main-belt asteroids in which strong radial color
gradients exist.  We find no statistically significant evidence for
bimodal or other non-uniform color distributions, either in our data,
or in data previously reported to show such evidence.  The impact
resurfacing hypothesis is re-examined in the light of the new color
data and is rejected as the primary cause of the observed color
dispersion.  We also present new near-infrared reflection spectra of
1993 SC, 1996 TS$_{66}$, 1999 DE${_9}$ and 2000 EB$_{173}$, taken at
the Keck and Subaru telescopes.  These spectra, combined with others
from the published literature, provide independent evidence for
compositional diversity in the Kuiper Belt.  Objects 2000 EB$_{173}$,
1993 SC, and 1996 TS$_{66}$ are spectrally bland while 1999 DE${_9}$
shows solid-state absorption bands.

\end{abstract}


\newpage

\section{Introduction}

Ground-based surveys have revealed (as of July 2001) more than 400
Kuiper Belt Objects (KBOs).  These bodies have orbital semi-major axes
$a > 30$ AU and a wide range of orbital properties showing clear
evidence for dynamical sub-structuring (Jewitt, Luu and Chen 1996; Luu
et al.  1997; Jewitt, Luu and Trujillo 1998; Trujillo, Jewitt and Luu
2000).  The inferred population of objects with diameters $\ge 100$ km
is of order $4 \times 10^4$ (in the classical belt; Trujillo, Jewitt
and Luu 2001) while the inferred number of (mostly unseen) bodies
larger than 1 km in diameter numbers many billions.

The Kuiper Belt is thought to be a relic from the earliest phases of
the solar system. The low radiation equilibrium temperatures of KBOs ($T \approx
40$ - 50 K) allow the possibility that many common ices could be
retained over the age of the solar system.  Indeed, it is widely
thought that the volatile-rich nuclei of most short-period
(specifically, Jupiter-family) comets are escaped members of the Kuiper
Belt while the Centaurs (bodies with perihelia and semi-major axes
between the orbits of Jupiter and Neptune) represent an intermediate
dynamical stage.  There is strong interest in understanding the
physical properties and compositional natures of the KBOs both as clues
to the early solar system and as samples of cometary material
relatively unchanged by heating by the sun.  Unfortunately, most known
KBOs are faint (apparent red magnitude $m_R \ge 22$) and their
properties difficult to measure with accuracy.  The most secure and, so
far, uncontested observational result is that the KBOs exhibit diverse
optical colors ranging from nearly neutral (V-R $\approx$ 0.3) to very
red (V-R $\approx$ 0.7 - 0.8) (Luu and Jewitt 1996; Green {\it et al.}
1997; Tegler and Romanishin 1998; Barucci {\it et al.} 2000).  A wide
range in the V-J color index also appears well established (Jewitt and
Luu 1998; Davies {\it et al.} 2000; Noll, Luu and Gilmore 2000).

Two qualitative explanations for the color diversity of KBOs have been
proposed (Luu and Jewitt 1996).  First, color differences might result
from the competition between cosmic ray bombardment (which may tend to
produce a low albedo, reddish reflectivity) and occasional impact
excavation of sub-surface volatiles.  The resurfacing hypothesis makes
specific predictions about the color distribution of the KBOs and is
therefore testable.  Resurfacing is only effective as a cause of
spectral diversity when the timescales for irradiation darkening and
impact resurfacing are comparable.  The resulting surfaces of the KBOs
should occupy a continuum of states from fully mantled, low albedo and
reddish at one end to fully resurfaced, ice covered and neutral at the
other. Furthermore, since not all resurfacing events would globally
coat KBOs with excavated material, a second prediction is that the
colors (and spectra) of KBOs should be azimuthally variable.  A third
prediction is that the rotationally averaged colors of the KBOs should
systematically vary with the object size, since small bodies have shorter resurfacing
times than larger ones while the cosmic ray reddening timescale is
independent of size (c.f. Figure 4 of Luu and Jewitt 1996).

Alternatively, color diversity could simply indicate intrinsically
different compositions among the KBOs.  This second hypothesis is
difficult to reconcile with the relatively small temperature gradients
expected across the trans-Neptunian solar system. In radiative
equilibrium with the sun, the local blackbody temperature varies as
$T(R) \approx 278 R^{-1/2}$, where $R$ [AU] is the heliocentric
distance.  The temperature difference from 30 AU to 50 AU is a paltry
11 K, scarcely enough to affect the chemical composition of the KBOs.
Slightly larger temperature gradients might be expected if the KBOs
have been swept outward from their formation locations by the radial
migration of Neptune, but still it is difficult to see how the bulk
composition could be substantially affected.  Nevertheless, in view of
the uncertainties in the modes of formation and dynamical histories of
the KBOs, the second hypothesis cannot yet be ruled out as a viable
explanation of the color distributions of the KBOs.

Tegler and Romanishin (1998 and 2000, hereafter TR98 and TR00) reported
a bimodal distribution of KBO colors.  Independently, Barucci {\it et
al.} (2001) refute this bimodality but report evidence for four
distinct color groups within the KBO population. Bimodal and
multi-modal distributions are interesting in the present context
because they cannot be produced by stochastic resurfacing.   The
question of whether the KBO color distribution is unimodal or bimodal
therefore confronts the resurfacing hypothesis.  While a number of
authors have published color measurements that specifically refute the
bimodal distribution (Barucci {\it et al.} 2000, Davies {\it et al.}
2000), it is in fact difficult to imagine how such a distribution could
be produced as an artifact.  For example, low quality photometry having
large uncertainties
 will tend to blur a bimodal distribution into a unimodal one, not the
other way round.  This raises the possibility that Tegler and
Romanishin's careful use of aperture correction techniques in
photometry could be responsible for their detecting bimodality while
others do not see it (Barucci $\it et al$.  2000 used aperture
correction but their sample included only eight objects).  We
investigate this possibility in Section 2.

In this paper, we present new optical photometry of 26 KBOs and 2
Centaurs taken at the Keck and University of Hawaii telescopes.  The
new measurements constitute a single data set relatively free of
systematic errors that might afflict conclusions drawn by comparing
photometry from different sources.  In addition, we present new
reflection spectra of four KBOs in the near infrared ($1 \micron \le
\lambda \le 2.4 \micron$) wavelength range.  We ask the following questions:

\begin{enumerate}
\item Are the optical colors bimodally distributed?

\item Are the optical and optical-near-infrared colors correlated?

\item Are there correlations between surface color and physical or orbital 
properties of the KBOs?

\item Are the available data compatible with the resurfacing hypothesis?

\item What clues are provided by the spectral evidence?
\end{enumerate}

\section{Optical Observations}

Optical observations were obtained under photometric conditions at the
Keck II telescope using the LRIS imaging camera (Oke {\it et al.} 1995).
This camera incorporates a Tektronix 2048 x 2048 pixel charge-coupled
device (CCD) and gives an image scale of 0.215 arcsec per $24\mu m$ pixel.
Facility broadband BVRI filters were used for all observations.  
Photometric calibration of the data was obtained through
observations of standard stars on the Johnson-Kron-Cousins photometric
system (Landolt 1992).  The images were processed using flat fields
calculated from the nightly medians of bias-subtracted data taken
through each filter.  The resulting images were found to be uniform in
sensitivity across the width of the CCD to better than 1\%, with the
largest deviations occurring near the east and west edges of the field
of view, in a region that was partially vignetted by the LRIS optics.
All target objects and standard stars were placed within 300 pixels of
the center of the field of view (FOV) and completely avoided the 
vignetting.  The images were autoguided at sidereal rates.  We used
short integration times (from 200 sec to 500 sec) in order to minimise
trailing of the KBOs relative to the stars.  At opposition, the KBOs
trail $\sim 0.4$ arcsec in 500 sec, which is small compared to the 0.7 - 1.0 arcsec
Full Width at Half Maximum (FWHM) seeing in most of our data.  

Additional optical observations of four bright KBOs were obtained at
the University of Hawaii 2.2-meter telescope.  We used a Tektronix
2048 $\times$ 2048 pixel CCD camera at the $f$/10 Cassegrain focus to obtain a
$7.5 \times 7.5$ arcmin$^2$ FOV and a 0.219 arcsecond/pixel image
scale. Flat fielding and photometric calibration were obtained as at
the Keck telescope.

\subsection{Profile Correction}

There are several sources of uncertainty in the photometry and, in view
of the importance of the errors, it seems worthwhile to discuss them
here.  The main problem is that, while large photometry apertures are
needed to capture all the light in the point spread function (PSF),
large apertures also capture a large (and noisy) signal from the
background sky.  Specifically, if the function $f(a)$ ($W m^{-2}
Hz^{-1} (arcsec)^{-2}$) represents the PSF, the total flux within the
PSF measured out to radius $a$ [arcsec] is simply

\begin{equation}F(a) = \int^a_0 2\pi af(a)da,\end{equation}

while the flux in the background measured out to the same
radius is

\begin{equation}F_b(a) = \int^a_0 2\pi af_bda = \pi a^2f_b,\end{equation}

\noindent where $f_b$ ($W m^{-2} Hz^{-1} (arcsec)^{-2}$) is the sky
surface brightness. Ideally, we would like to measure $F(a \rightarrow
\infty)$, to capture all of the light from the PSF but then background
signal $F_b(a \rightarrow \infty) \rightarrow \infty$. For faint
objects, the noise on the background sky completely dominates the
photometric uncertainty.  The tempting option is to use photometry from
small apertures to reduce random errors caused by noise on the
background signal. However, this introduces potentially devastating
systematic errors due to the variable and unknown fraction of the total
light captured within the small aperture.

We followed the example of TR98 and used photometry of field stars to
empirically determine a profile correction, defined by

\begin{equation}\Delta m_p = -2.5log\left[\frac{F(a_1)}{F(a_2)}\right].\end{equation}

Here, $a_1$ and $a_2$ are the radii of the small and large photometry
apertures used, respectively, to measure the KBOs and their reference
field stars and the photometric standard stars.  Essentially, Eq. (3)
is a correction to be added to small-aperture measurements of KBO
magnitudes to estimate the total flux density that would be obtained if
low noise, large aperture measurements were possible.  We experimented
to determine optimal values of the aperture radii by trial and error,
eventually selecting $a_1 = 1.00$ arcsec and $a_2 = 3.22$ arcsec for
most measurements.  The median sky signal was determined from a
contiguous annulus 4 to 5 arcsec wide.  Typically 5 - 10 field stars
were identified in each image and used to determine $\Delta m_p$.  In
practice, we found two important limitations to this procedure.  First,
the PSF of the Keck data varies slightly with position on the CCD.  We
selected field stars near the target KBOs to reduce the effect of
spatial variations.  Second, in the fine seeing above Mauna Kea we
could see that most of the field objects were marginally resolved
galaxies and, in some fields, we had difficulty locating a suitable
number of stars with which to determine $\Delta m_p$.  With the above
parameters, we found $\Delta m_p = 0.3$ to 0.5 mag, depending on the
seeing.  The uncertainty on $\Delta m_p$ determined from the dispersion
among measurements of different field stars was typically $\sigma_p$ =
0.01 - 0.03 mag.

\subsection{Noise Properties of the Data}

We examined the effects of background fluctuations on the photometry by
placing artificial KBOs on the CCD and measuring their magnitudes using
procedures identical to those employed for real objects.  The PSFs of
the artificial KBOs were matched to the PSFs of the real data on an
image-by-image basis, including allowance for non-zero ellipticity
(e.g., caused by wind shake) where appropriate.  Measurements of about
20 artificial objects were used to calculate the empirical standard
deviation within each of several magnitude bins.  A sample image is
shown in Fig. 1, for the particular case of 1996 SZ4.  In
background-limited observations, the photometric uncertainty,
$\sigma_B$, should vary with the magnitude of the source, $m_R$, as

\begin{equation}\sigma_B = -2.5\log\left[1 + \kappa 10^{0.4m_R}\right]\end{equation}

\noindent where $\kappa$ is a constant that depends on the surface
brightness of the night sky and the integration time.  We find that
Equation (4) provides an acceptable fit to the measurements in each
filter (Fig. 2).  The combined uncertainty is estimated from

\begin{equation}\sigma = \sqrt{(\sigma_B^2 + \sigma_p^2)}.\end{equation}

The above error treatment accounts for contamination by background
sources in a statistical sense.  The photometry can potentially still
suffer from contamination by rare, bright background sources that are
not well sampled by the above procedure.  Most such sources are distant
galaxies, typically extended on scales comparable to the seeing. For
example, a galaxy of red magnitude $m_R = 25$ constitutes a $2.5\%$
($6.5\%$) error signal to a foreground KBO of magnitude $m_R = 21
(22)$.  We used the motion of KBOs between images to provide protection
from such background objects.  Photometry that was obviously
compromised was rejected from the sample.

\subsection{Photometric Results}

The geometrical circumstances of the observations are summarised in
Table 1.  The optical photometry is listed in Table 2 while color-color
plots are shown in Figures 3, 4 and 5.  For reference, the
corresponding colors of the Sun are approximately given by $B-V$ =
0.67, $V-R$= 0.36, $R-I$ = 0.35, $V-J$ = 1.08 (Hartmann, Cruikshank and
Degewij 1982, Hartmann {\it et al.} 1990).  The solar colors are marked
in Figures 3 - 5. We note immediately from the Figures that the B-V, V-R
and R-I color indices are redder than the Sun and are mutually
correlated:  objects that are red at B (4500\AA\ wavelength) are also red
at I (8500\AA).  Linear correlation coefficients are summarized in
Table 3.  In the Table, $P(r \ge r_{corr})$ denotes the probability
that a correlation coefficient equal to or larger than the one measured
could be produced by chance in random, uncorrelated data.  $P(r \ge
r_{corr}) = 0.003$ corresponds to the nominal $3\sigma$ criterion for a
statistically significant correlation.

The color-color correlations are extended to longer wavelengths in Fig.
6, where we plot those objects observed in J band ($1.2 \micron$) from
Jewitt and Luu (1998) and Davies {\it et al.} (2000).  The B-I and V-J
color indices are highly correlated (Table 3), suggesting that optical
colors may be used as a proxy for the optical-near infrared V-J color.
The correlations show that the reflectivities of the KBOs are
consistent with a single reddening agent in the B ($0.45 \micron$) to J
($1.2 \micron$) regime.  Spectra of many low albedo hydrocarbons show
exactly this property (Cloutis {\it et al.} 1994).  The reflectivity
gradients at wavelengths greater than $1.2 \micron$ tend to be smaller,
and are not well correlated with the optical gradients (Jewitt and Luu
(1998), Davies {\it et al.} (2000)).  A subset of the complex
hydrocarbons also mimic this property, notably some coal tar extracts
studied by Cloutis {\it et al.} (1994) and the Tholins (Cruikshank {\it
et al.} 1998).  The colors thus suggest, but do not uniquely diagnose,
the presence of surface hydrocarbons.

The mean optical colors of the classical and resonant KBOs are
indistinguishable (Table 4).  The optical colors are correlated with
neither the orbital nor physical properties of the KBOs (Table 3).  If
collisional effects are important, we might expect a correlation
between KBO color and $\Delta V$, the difference between the average
velocity and the velocity of an uninclined, circular orbit of the same
semi-major axis.  We computed $\Delta V$ from the eccentricity and
inclination of each of the observed KBOs, but find no statistically
significant correlation with the $B-I$ color (Table 3).  A V-J vs.
semi-major axis correlation was tentatively claimed by Weintraub {\it
et al.} (1997) based on measurements of four Centaurs.  In fact, this
correlation is statistically insignificant ($P(N,r_{corr}) < 0.1$) and
is unsupported by the new data (Table 3).

Some of the objects in our sample have been previously measured.  We
compare the new results with published data in Table 5.  A graphical
comparison of the measurements is shown in Figure 7. In general, the
independent measurements agree within the combined uncertainties.  For
example, 5 of 19 objects (26\%) have independently measured $V-R$
colors which are different by more than 1 combined standard deviation
($1\sigma$).  This is consistent with a Gaussian error distribution,
for which we expect about 32\% of measurements to lie beyond
$1\sigma$.  We summarise the data sets in Table 6.  In terms of their extrema,
means and dispersions, the color samples of TR98, TR00 and the present
work are remarkably concordant (columns 3-5 in Table 6).

\subsection{The Color Distribution}

TR98 contend that the B-V and V-R colors are bimodally distributed.
Their data, which include a mixture of Centaurs and KBOs, are plotted
in Fig. 8.  In a later work, Tegler and Romanishin 2000 reported
observations of the B-V and V-R colors of 16 additional KBOs and 3
Centaurs (Fig. 9).  They reported that the color distribution remained
bimodal with the addition of the new data, although this is visually
less obvious (compare Fig. 8 with Fig. 9).  Tegler and Romanishin did
not assess the statistical significance of the bimodality in either of
their samples, and neither did they attempt to account for its physical
origin.  Clearly, a test of the significance of the reported bimodality
is in order.  We consider three such tests. \\

\noindent {\bf The Bin Test}

Most of the dispersion among the colors in Figs. 8 and 9 occurs along a
line of principal variation that is slightly inclined relative to the
B-V axis.  We use the position of each KBO measured along this line,
defined by

\begin{equation}C = \left[(B-V)^2 + (V-R)^2\right]^{1/2}\end{equation}

\noindent to test the null hypothesis that the colors are distributed
randomly along the line.  The probability that, in a sample of $n$
measurements distributed randomly among $k$ equal-sized bins, the
central bin will hold $m$ objects is given by

\begin{equation}P(n,m) = \left[\frac{n!}{m!(n-m)!}\right]\ p^m (1-p)^{n-m}\end{equation}

\noindent where $p = 1/k$.  We present calculations for three bins ($k
= 3$) with $C$ in the range $0.65 \le C \le 1.55$.  Table 6 lists the
number of objects in the central bin, which has $0.95 \le C \le 1.25$,
for each sample. 

Evaluation of the probabilities using Eq. (7) shows that the TR98
sample has roughly a $3\%$ likelihood of having been drawn by chance
from a uniform distribution, corresponding to a result that is $\approx
2.2\sigma$ in statistical significance.  Tegler and Romanishin's larger
second sample (TR2000) is even less significantly bimodal (with a
$13\%$ probability of having been derived from a uniform distribution),
as is the sample from the present work (for which the probability is
also about $13\%$:  Table 6).  In fact, the combined TR98 + TR00 data
sets are comparable to the data set from the present work in terms of
size, range, mean, standard deviation on the mean, and the lack of
evidence for a bimodal distribution of colors (Table 6).  In all, we
find that the data present a remarkably coherent case for a
distribution of colors that is devoid of evidence for bimodality.

The bin test has been recomputed for other values of $C_{min}$,
$C_{max}$ and $k$, but always with the results that a) the most
significant (but still $< 3\sigma$) evidence for bimodality is in the
TR98 data and b) the significance of bimodality decreases as the sample
size increases.  We conclude that the bin test provides no evidence
that the $B-V$ or $V-R$ colors are distributed bimodally.\\

\noindent {\bf The Dip Test}

As another test for bimodality, we make use of the dip statistic
(Hartigan and Hartigan 1985), which is defined as the maximum
difference between the (perhaps bimodal) data distribution and the
unimodal distribution function that minimizes that maximum difference.
In effect, the test tries to explain the data distribution with a
best-fitting unimodal distribution, and the dip statistic is the
residual from matching the two functions.  The larger the dip the
larger the mismatch.  Hartigan and Hartigan's test computes the dip
statistic then evaluates its significance. The results upon performing
the dip test on Tegler and Romanishin's data and our own data (this
work) are presented in Table 7. In agreement with visual impressions
(Figure 8), the most nearly significant evidence for bimodality is in
$V-R$ from TR98 but this is still $< 3\sigma$ and not supported by the
TR00 or present $V-R$ measurements.  The dip test, like the bin test,
provides no evidence in support of bimodality.  \\

\noindent {\bf The Interval Distribution Test}

We lastly consider the distribution of intervals between KBO color
measurements. In a truly bimodal distribution, there should be a large
interval (between modes) followed by many smaller intervals (between
the members of each mode).  Conversely, in a continuous distribution,
one expects all intervals to be roughly equal.  To assess the
likelihood that the large interval observed in Tegler and Romanishin's
color distribution might arise by chance from random, uncorrelated
data, we computed Monte Carlo models in which we randomly picked colors
from a uniform distribution, then determined the largest interval, $LI$,
between two consecutive colors.  We repeated this experiment
$10^5$ times in order to determine the probability of obtaining a given
$LI$ by random chance, given $n$ data points.  We selected colors in
the ranges $0.60 \le B-V \le 1.30$ and $0.35 \le V-R \le 0.80$,
respectively.  For illustration, we present the results of our
experiments in Figs. 10 and 11, where we plot the $LI$ cumulative probability
distribution as a function of $n$ for the $B-V$ and $V-R$ data sets.

The results of the interval distribution test are summarised in Table
8.  There it can be seen that, at the canonical $3\sigma$ ($P = 0.3\%$)
level of confidence, the measured values of $LI$ are consistent with
random sampling of a uniform color distribution.   This is already
qualitatively obvious in a comparison of Fig. 8 (TR98) with Fig. 9
(TR00).  The gap in Fig. 9 is partially filled-in relative to the gap
in Fig. 8, just as one expects from an enlarged data sample if the
gap is merely a statistical fluctuation.  A real gap in the color
distribution would not shrink as the sample size increases.

We conclude that both our photometry and the photometry presented in
two papers by Tegler and Romanishin are formally consistent with
derivation from a uniform distribution of colors.  Similarly, the
distribution of optical-infrared ($V-J$) colors provides no hint of a
bimodal distribution (Davies $\it et al.$ 2000).  These results are
compatible with the original finding of Luu and Jewitt (1996) and with
subsequent results presented by independent workers (Green {\it et al.}
1997, Barucci {\it et al.} 2000).  Finding no significant evidence for 
bimodality, we next re-consider impact resurfacing as a possible
cause of the observed unimodal color distribution.

\section{Resurfacing Revisited}

We have revisited the resurfacing mechanism taking into account
improvements in the known parameters of the KBO size distribution that
have accrued from our Mauna Kea surveys (Jewitt, Luu and Trujillo 1998,
Trujillo, Jewitt and Luu 2001).  The most important change that has
occurred since Luu and Jewitt (1996) is the improved determination of
the size distribution index.  The Kuiper Belt Objects are distributed
such that the number of objects with radii in the range $a$ to $a + da$
is $n(a)da = \Gamma a^{-q}da$, with $\Gamma$ and $q$ constant.  The
best-fit index $q = 4.0^{+0.6}_{-0.5}$ (Trujillo, Jewitt and Luu 2001)
is larger than the steepest models considered in Luu and Jewitt
(1996).  The total number of classical KBOs (CKBOs) larger than 100 km
in diameter remains unchanged from our previous estimates,
at $N_{CKBO} = 3.8^{+2.0}_{-1.5}$ x $10^4$ (Trujillo, Jewitt and Luu
2001).

The collision rate onto a target KBO of radius $a_T$ is

\begin{equation}  1/\tau \approx \int^\infty_{a_{min}} 4\pi (a_T + a)^2 \Delta V \psi n(a)da /W, \end{equation}

where $\Delta V$ is the velocity difference between the target and
incoming projectiles, $\psi$ is a dimensionless factor for
gravitational focussing and $W$ is the volume swept out by the orbits
of the KBOs.  Here, $a_{min}$ is the radius of the smallest
projectile.  In practice, the integration extends to the largest KBO in
the distribution (not to infinity!) but, given the steep size
distribution, impacts by the largest objects are comparatively rare and
the integration upper limit is not critical.  We take $\Delta V =$ 1.3
$km$ $ s^{-1}$ (Trujillo, Jewitt and Luu 2001).  The gravitational 
focussing is given by $\psi = 1 + (v_e/\Delta V)^2$, where $v_e$ is the
escape speed from the target object.  For a spherical KBO of density
$\rho$ $[kg$ $ m^{-3}]$, this may be expressed as

\begin{equation}\psi = 1 + 8\pi G \rho a_T^2 / (3 \Delta V^2)\end{equation}

Substituting $\Delta V$ = 1.3 $km$ $s^{-1}$, $\rho$ = $10^3$ $kg$ $m^{-3}$,
we see that $\psi$ differs from unity by no more than $10\%$ for 
$a_T \le 500$ $km$, meaning that we can safely set $\psi$ = 1 for all
the objects considered here.  The classical Kuiper
Belt is well represented by an annulus with inner and outer radii $R_i
= 30 AU$, $R_o = 50 AU$, respectively, and thickness $H = 10 AU$, for
which the volume is $W$ = $\pi H (R_o^2 - R_i^2)$ = 2 x $10^{38} m^3$.
With these parameters and with $q =$ 4, $a_{min}$ = 1 $km$ and $a_T$ = 50 $km$ we
obtain $1/\tau \approx 10^{-2} Myr^{-1}$ for objects in the classical
Kuiper Belt.  The rate increases with decreasing projectile size (for
example, with $a_{min}$ = 0.1 $km$, the corresponding rate is $1/\tau
\approx 10^{1} Myr^{-1}$), showing the importance of small impacts.

For the resurfacing model, we follow the prescription in Luu and Jewitt
(1996).  From Equations 5-7 and 9 of that paper, the ejecta blanket
radius is found to be

\begin{equation} r_{eb} = 17 km (a/1km)^{0.69} / (a_T/100km)^{0.31}. \end{equation} 

We again assume $q = 4.0$, noting that this index is determined for large
KBOs ($a > 50$ km) and may not hold for the smaller objects which
dominate collisional resurfacing.  We computed Monte Carlo simulations
for 1 $Gyr$ at a time resolution of 1 $Myr$, using a 1000 x 1000 pixel
array to represent the surface.  Other important parameters include the
timescale for radiation damage of the surface, which we take to be
$10^8$ yrs (Luu and Jewitt 1996) and the minimum impactor size, for
which we assumed a range of values $0.05 \le a_{min} \le 1.0$ km.  
Sample models with $a_{min}$ = 0.05 km are shown in Figure 12.

The principal features of the data that we seek to compare with the
resurfacing simulations are 1) the wide spread in the surface colors
and 2) the absence of a measurable color-diameter trend.  As noted in
Luu and Jewitt (1996), steeper projectile size distributions increase
the weight given to numerous, small impacts relative to rare, global
resurfacing events.  For this reason, the stochastic character of the
new simulations is smaller than in Luu and Jewitt 1996.  In fact, the
$q = 4$, $a_T$ = 50 km models show much less scatter than do the real
KBOs (Figure 12).  In the observational sample the standard deviation
on the mean of 28 measurements is $\sigma(V-R)$ = 0.02 mag (Table 4),
corresponding to 0.11 mag. per object.  The peak-to-peak color range in
$V-R$ is 0.5 mag. (Figure 3). On the other hand, the resurfacing models
typically give much more muted variations with peak-to-peak 
excursions $<$ 0.1 mag.  Our first conclusion is that the color
dispersion observed among the KBOs is larger than given by
our $q = 4$ resurfacing model.

The resurfacing simulations predict a color-diameter trend, caused by
the size dependence of the impact resurfacing time relative to the
(constant) timescale for cosmic ray irradiation damage of the surface
(Figure 12).  The small KBOs are rapidly and repeatedly resurfaced,
preventing them from reddening under the effects of cosmic rays whereas
the large KBOs are difficult to resurface and therefore more susceptible to
cosmic ray irradiation.  Small KBOs also display substantial color
dispersion relative to the larger objects because the colors are less
affected by spatial averaging.  However, the color differences as a
function of KBO size are modest.  In the size domain of the observed objects,
between $a_T$ = 50 km and $a_T$ = 250
km, the mean $V-R$ varies by only 0.1 mag. (Figure 12).  A factor of 5
in radius corresponds, at constant albedo, to a 3.5 magnitude
brightness difference.  Inspection of the data shows that a
$0.1/3.5$ [mag./mag.]
color-magnitude gradient is too small to be seen in the real KBO
sample, in view of the large intrinsic scatter.  We conclude that the
absence of a color-diameter trend in the KBO sample does not by itself
constitute evidence against the action of resurfacing.

Perhaps the least model-dependent and, therefore, most serious blow
against the resurfacing hypothesis is its prediction that, for objects
of a given diameter, azimuthal color variations should be equal in
magnitude to the color dispersion among objects.  That this is not the
case among the real KBOs may be seen by inspection of Table 5.  There,
repeated determinations of the colors taken at random rotational phases
are in general agreement within the quoted uncertainties of measurement
while color differences between the KBOs are many times larger.  The
most significant exception is 1996 TL$_{66}$.  Even here, however, the
$V-R$ color range of $0.24 \pm 0.08$ mag.  is small compared to the 0.5
mag. peak-to-peak color differences between objects (Tables 2, 3).  If
resurfacing were the primary cause of the well-established color
differences between KBOs, then rotational color variations from $V-R$ =
0.35 to $V-R$ = 0.80 would be seen in $\it individual$ KBOs, and this
is clearly not the case.

We conclude that impact resurfacing is probably not the primary cause
of the color differences which exist among the KBOs.  It remains
possible that resurfacing plays a secondary role, and may be
responsible for azimuthal spectral variations like those reported on
Centaur (8405) Asbolus (Kern et al. 2000).   As noted in Luu and Jewitt
(1996), resurfacing produces a range of colors only if the timescales
for impact resurfacing and cosmic ray reprocessing are of the same
order.  Neither timescale is easy to estimate. The irradiation
timescale depends on the cross-section for interaction between the
cosmic rays and the molecules comprising the KBO surface layers.  This
cross-section is highly composition dependent.  The resurfacing
timescale is dominated by the abundance of impactors which are too
small to be detected in current surveys and whose statistics are
therefore uncertain. Furthermore, impact gardening of KBO regoliths may
be important in exposing previously buried material, as on the Moon, and
modelling this effect introduces additional levels of arbitrariness.
The net result is that we cannot argue, on physical grounds, that the
relevant timescales are of the same order of magnitude, as is required
for the hypothesis to succeed.
Neither is it completely obvious that the color of an irradiated ice
mixture changes with the fluence in the simple way assumed by the
resurfacing hypothesis.  Indeed, laboratory experiments show a more 
complicated relationship.

It is unlikely that simple grain size effects could be responsible for
such a wide color diversity (Moroz $\it et al.$ 1998).  Instead, real
compositional variations seem to be responsible.  In the main (Mars -
Jupiter) asteroid belt, some colorimetric diversity is produced by
impact shattering of internally differentiated precursor bodies.  This
is not likely to be the explanation in the Kuiper Belt, where the 100
km and larger objects studied here are too large to have been produced
by fragmentation of precursors (Farinella and Davis 1996).  As noted
above, the compositional gradient in the main-belt caused by a strong
impressed radial temperature gradient is also unlikely to be relevant
in the Kuiper Belt, where temperatures are low and temperature
differences are small.  The primary origin of the color dispersion
remains unknown.

\subsection{Other Reported Correlations}

Tegler and Romanishin (2000) reported that all 9 observed KBOs with
perihelia $q > 40$ AU systematically belong to their "red" group, while
objects with smaller $q$ are distributed more evenly between the red
and grey groups.  They showed that the probability of finding this
result by random selection from a bimodal distribution is about
$1/512$, corresponding to about $3\sigma$ in a Gaussian error
distribution and argued that the effect is therefore of probable
statistical significance.

We have re-examined Tegler and Romanishin's 1998 and 2000 data without
making the classification of objects into "red" and "grey", since the
data provide no support for this division.  We use the $B-R$ color
index as our metric, as this is more robust relative to measurement
errors than either $B-V$ or $V-R$ separately.  In their combined sample
of 32 objects having $B-R$ color indices, there are 26 with $q <$ 40
AU, for which the median value $(B-R)_m$ = 1.54.  All 6 objects with $q
>$ 40 AU have $B-R > (B-R)_m$.  Given that the probability of any
measurement falling above the median is, by definition, $1/2$, the
chance that all of 6 values would be larger than $(B-R)_m$ is
$(1/2)^{6}$ = 1/64.  The combined data sets from TR98, TR00 and this
paper (Table 2) include 45 different objects with measured $B-R$
indices. We averaged separate measurements of multiply observed objects
to obtain one $B-R$ per object.  The median color for the 36 objects
with $q <$ 40 AU is $(B-R)_m$ = 1.46, while all 9 objects with $q >$ 40
AU have $B-R > (B-R)_m$.  If the $q > $ 40 AU objects have the same
median color as those with smaller perihelia, the probability of
obtaining the observed result is $(1/2)^{9}$ = 1/512.  We therefore
concur with TR00 that there is statistical ($\approx 3\sigma$) evidence
for a color difference between objects having perihelia on either side
of the $q =$ 40 AU boundary.  Clearly, however, this remains a weak
result that should be tested by substantially increasing the sample
size, particularly for objects with $q >$ 40 AU.  Until this is done,
we see little point in speculating about possible causes of the color
difference.

Levison and Stern (2001) reported a size dependence of the inclination
distribution of KBOs, such that objects with absolute magnitude, $H <
6.5$ have a wider inclination distribution than others.  However, using
the Kolmogorov-Smirnov test, they find a $3\%$ probability that the
measured $H < 6.5$ and $H > 6.5$ distributions are drawn from the same
parent population, corresponding to a result with $97\%$ confidence.
This, in turn, corresponds to about $2.2\sigma$ in a Gaussian probability
distribution and is therefore formally insignificant.  Our own data
show no evidence for a color - magnitude trend (Table 3), and no evidence
that the $H < 6.5$ and $H > 6.5$ color distributions are different.

\section{Near-Infrared Spectra}

Near infrared spectral observations were obtained using the Keck I 10m
and the Subaru 8-m telescopes.  At Keck, we used the NIRC spectrometer
(Matthews and Soifer 1994) at the $f$/25 forward Cassegrain focus.
NIRC contains a $256 \times 256$ pixel InSb array with $30 \micron$
pixels and a corresponding image scale of 0.15 arcsec per pixel (38
arcsec FOV).  A slit of projected dimensions 0.68 arcsec $\times$ 38
arcsec was used in conjunction with 150 line/mm and 120 line/mm grisms
for the JH and HK spectral regions, respectively, giving spectral
resolutions $ \lambda/ \Delta \lambda \approx 100$.  Observations were
taken in several steps.  First, NIRC broadband images were used to
identify the target KBOs by their motion relative to the fixed stars.
Second, non-sidereal rates were entered into the telescope control
system in order to follow the motion of the KBO.  The telescope was
moved while autoguiding to place the object at the slit location and
the alignment was checked, iteratively, using further broadband
images.  Next, the slit and grism were inserted into the beam.  A
spectrum was taken while dithering the target image between two
positions along the slit separated by 10 arcsec.  The alignment was
re-checked every 20 - 30 minutes by removing the slit and re-imaging
the target.  Spectral flat fields were obtained by imaging a diffusely
illuminated spot inside the dome.  Spectral calibration of the images
was obtained from nearby stars on the UKIRT Faint Standards list (Hawarden
$\it et al.$ 2001).  In addition, we observed nearby solar analogue stars
in order to cancel features specific to the solar spectrum.

At the Subaru 8-meter telescope we employed the CISCO camera in
spectroscopic mode to acquire separate spectra in the JH and HK
windows.  CISCO is a high-throughput grating-dispersed near infrared
spectrometer using a $1024 \times 1024$ HgCdTe array of $18.5 \micron$
pixels as detector (Motohara {\it et al.} 1998).  The pixel scale is
0.11 arcsec/pixel, giving a 110 arcsec $\times 110$ arcsec FOV.  The
spectral resolution was $\lambda / \Delta \lambda = 1000$.  We
identified the targets again through their proper motions.  At the time
of use, Subaru could not track at non-sidereal rates.  For this reason,
we observed 2000 EB$_{173}$ near its stationary point to minimise the
angular motion.  As precautions, we used a relatively wide (1.0 arcsec)
slit and we aligned the slit parallel to the apparent proper motion
vector so that the natural motion of the object would carry it along,
rather than out of the slit.  As with NIRC, the alignment was
periodically checked using broadband images and corrected when
necessary.  Flat fields were created from medians of the data.
Calibration was again obtained using stars from Hawarden $\it et al.$ 2001.
The parameters of the spectral observations are listed in
Table 9.

Spectral data reduction included steps to create and apply a bad pixel
mask, flattening of the data, subtraction of night sky lines using
adjacent dithered image pairs, removal of night sky line residuals by
interpolation through the object position and, finally, extraction of
the object spectrum.  Variability in the transmission of the
atmosphere, particularly in the water bands near $1.9 \micron$, also
limit the photometric accuracy in some of the data.  Fortunately, most
observations were taken in dry atmospheric conditions and it was
possible to take useful data through the atmospheric water absorption
band. Nevertheless, systematic errors due to night sky lines and
variable atmospheric extinction rival or dominate statistical errors in
the spectra from NIRC.  The higher resolution of the Subaru CISCO
spectra allowed more accurate subtraction of the night sky lines.  The
signal per resolution element in the CISCO data is very small, however,
so that we must bin the data to obtain useful spectral information.

The reduced spectra are shown in Figs. 13-17, where they have been
divided by the spectrum of the Sun and normalized to unity at $2.2
\micron$.  In Table 10 we list the normalized reflectivities binned to
$0.1 \micron$ resolution.  Error bars on the binned points are
$1\sigma$ standard deviations on the means of the pixels within each
bin.  For each spectrum in Figs. 13-17 we show the individual data
points and overplot the binned data.  We have included the spectrum of
1996 TL$_{66}$ from Luu and Jewitt (1998) for comparison with the new
data.  It was obtained at the Keck telescope using NIRC and observing
procedures identical to those employed for the other KBOs.

\subsection{2000 EB$_{173}$}

Meaningful interpretation of the spectra demands a full understanding
of systematic effects due to imperfect atmospheric extinction and sky
line cancellation.  These are best estimated empirically, by comparing
spectra of standard stars taken at different times and airmasses.  The
cleanest case is 2000 EB$_{173}$ (Fig. 13), for which the night was dry
and stable and the $1.9\micron$ telluric absorption feature is
correspondingly well cancelled.  The binned data suggest a broad
absorption band centered near $2.0 \micron$ but, in view of the likely
systematic uncertainties, we do not claim this feature as significant.
We agree with Brown {\it et al.} (2000), that the spectrum of 2000
EB$_{173}$ is featureless in the $1.0 \le \lambda \le 2.5 \micron$
wavelength range.  Unique compositional diagnoses of featureless
spectra are obviously impossible.  It is interesting to note, however,
that laboratory reflection spectra of highly carbonized materials are
commonly neutral and featureless, reflecting a deficiency of hydrogen bonds
(Cloutis {\it et al.} 1994, Moroz {\it et al.} 1998).  In the KBOs, we
expect that hydrogen will be mobilized in the surface layers by
continued cosmic ray bombardment, leading to eventual escape and a net
hydrogen depletion regardless of the initial composition.

\subsection{1999 DE$_9$}
In contrast, 1999 DE$_9$ is spectrally structured (Fig. 14).
Absorption features are evident near $1.4 \micron$, $1.6 \micron$,
$2.00 \micron$ and (possibly) at $2.25 \micron$.  The continuum drops
from $1.3 \micron$ down to the limit of the spectrum at $1.0 \micron$,
which we interpret as an additional broad absorption centered at or
shortward of $1.0 \micron$.  Water ice has well-known features at $1.55
\micron$, $1.65 \micron$, and $2.02 \micron$. The $1.55 \micron$ and
$1.65 \micron$ features appear very weakly in the DE$_9$ spectrum; we
note this is a characteristic of spectra of fine grain frost (Clark
1981a).  Water ice also has a high-overtone band at $1.25 \micron$,
which is not seen in the DE$_9$ spectrum.  The suppression of
high-overtone bands relative to the low-overtones (bands at longer
wavelengths) is observed when other minerals are present in addition to
water ice (Clark 1981a), and this might explain the absence of
the $1.25 \micron$ feature in 1999 DE$_9$. The strongest water band at
$2.00 \micron$ is only about 10\% deep.  The general weakness of the
water ice features suggests a low abundance of water ice, or water ice that
is heavily contaminated by an absorbing component. Indeed, the
1999 DE$_9$ spectrum is very similar to that of a mixture of Mauna Kea
red cinder and 1\% (by mass) water ice (Clark 1981b), shown in Figure
18.  Notice that the $\approx 1.6 \micron$ features in the cinder
spectrum are extremely weak, as in 1999 DE$_9$, and that the $\approx
2.0 \micron$ bands in the KBO and the Mauna Kea sample match in both
position and width.

Features near $1.4 \micron$ and $2.25 \micron$ are usually diagnostic
of the presence of metal-OH combination and overtone vibrational
motions in minerals that incorporate OH within their crystal structure
(Hunt 1977).  The Mauna Kea spectrum in Figure 18 shows absorption near
$1.4 \micron$ and a possible band at $2.22 - 2.25 \micron$.  The exact
location of these bands can be diagnostic of the compound (either Al or
Mg) associated with the OH stretch, but the quality of the DE9 spectrum
is not good enough for us to make this identification.  Given the
limited spectral coverage and the unknown albedo, we tentatively assert
that the spectrum of 1999 DE$_9$ shows evidence for the hydroxyl group,
with the latter interacting with an Al- or Mg-compound. If it can be
confirmed, the metal-hydroxyl identification would be most significant
because the reactions that create the metal-OH compounds proceed only
in the presence of liquid water, implying that temperatures near the
melting point have been sustained in 1999 DE$_9$ for at least a short
period of time.  Localised impacts on surface materials on 1999 DE$_9$
might provide a sufficient (although transient) heat source.  Heating
of core volatiles by the prolonged decay of radioactive elements could
also lead to melting but a mechanism of transport to the surface would
in addition be required for these materials to be spectrally
observable.

A similar absorption near $2.27 \micron$ is present in the spectrum of
Centaur (5145) Pholus (Davies, Sykes and Cruikshank 1993, Luu, Jewitt
and Cloutis 1994), with which 1999 DE$_9$ shares a resemblance in the
$1.4 < \lambda < 2.4 \micron$ wavelength range (Figure 19).  This
feature has been tentatively interpreted as due to solid methanol
(CH$_3$OH; Cruikshank {\it et al.} 1998).  However, as Cruikshank {\it
et al.} note, the identification with methanol is uncertain in part
because an expected second band at $2.33 \micron$ is not observed.  The
Pholus spectrum modelled by these authors has an unfortunate gap at
$1.4 \micron$. We recommend that this wavelength be observed to search
for evidence for the counterpart metal-OH absorption feature suspected in
1999 DE$_9$.

Lastly, our spectrum of 1999 DE$_9$ shows a broad absorption in the
$1.0 < \lambda <$ 1.3 $\micron$ wavelength region with an apparent
minimum near $1.05 \micron$.  Unfortunately, the minimum is close to
the short wavelength end of our spectrum and we cannot be confident of the
trend of the reflectivity at $\lambda \le 1.0 \micron$.  A true minimum
near this wavelength would be suggestive of the presence of a ferrous
silicate, possibly an olivine (Cloutis and Gaffey 1991).  
Forsterite (Mg$_2$SiO$_4$, a magnesium rich
olivine) has been independently proposed to fit a 1$\micron$
absorption in the reflection spectrum of Centaur (5145) Pholus
(Cruikshank {\it et al.} 1998) and would fit the $1\micron$ absorption in 
1999 DE$_9$. The identification of Forsterite as the
specific form of silicate is non-unique but it is attractive because of
the known presence of Forsterite in the dust ejected from comets
(Crovisier {\it et al.} 2000).  The single object 1999 DE$_9$ thus
provides plausible compositional ties to both the Centaurs and the
nuclei of comets (which are known to be water rich) through the water
ice (Foster {\it et al.} 1999, McBride {\it et al.} 1999, Luu, Jewitt
and Trujillo 2000) and silicate (Cruikshank {\it et al.} 1998) bands.
On the other hand, if the reflectivity continues to decline at $\lambda
\le 1.0 \micron$, we would instead suspect that the absorption is due to
the same complex hydrocarbons (e.g. Tholins) that give reddish optical
colors to the KBOs.  Only new spectra will allow us to decide between
these interpretations.

\subsection{1993 SC}

Numerous absorptions were reported in the near infrared spectrum of
1993 SC by Brown {\it et al.} (1997).  Their spectrum was taken at Keck
with the NIRC spectrometer over a range of airmasses from 1.2 to 1.7
and with an integration time of 3600 seconds.  They reported absorption
features at $1.62 \micron$, $1.79 \micron$, $1.95 \micron$, $2.20
\micron$ and $2.32 \micron$, with depths up to 50\% of the local
continuum.  The features were not precisely identified by Brown {\it et
al.}, but a broad classification as hydrocarbon absorptions was made.
We targetted the H-K spectral region of 1993 SC specifically to examine
the features reported by Brown {\it et al.} Our integration of 6000
seconds was recorded between airmasses 1.04 and 1.09, with sky
cancellation using a nearby star observed at airmass 1.08.

None of the features reported by Brown {\it et al.} are apparent in our
data.  Instead, using the same instrument on the same telescope but
with a longer integration time, we find only a noisy continuum devoid
of significant spectral structure (Fig. 16).  In our spectrum, the
reflectivity in the $2.25 \micron - 2.35 \micron$ bin is $1.01 \pm
0.15$. This is incompatible with Brown {\it et al.}'s deepest minimum
at $2.32 \micron$ at the $3.3\sigma$ level.  (Brown {\it et al.}
presented their data heavily smoothed by convolution with a Gaussian,
which impedes a more detailed comparison with the unsmoothed data in
Fig.  16).  Simple experiments in which our data were smoothed by
convolution with a Gaussian generated (apparently deep but unreal)
features with wavelengths that did not match those reported by Brown
{\it et al.} As another metric of comparison, we compute,
$S_{2.2}/S_{1.6}$, the ratio of the reflectivity at K band ($S_{2.2}$)
to that at H band ($S_{1.6}$).  In the Brown {\it et al.}  spectrum
this ratio is $S_{2.2}/S_{1.6}$ = 2.2 (see their Fig. 1), while we
measure $S_{2.2}/ S_{1.6} = 1.0 \pm 0.1$ (Fig. 16).  Finally, earlier
broadband photometry gives H-K = $-0.04 \pm 0.19$ (Jewitt and Luu 1998)
which, with the solar H-K = 0.06, corresponds to $S_{2.2}/ S_{1.6}$ =
$0.91\pm0.19$.  The filter photometry and our new spectrum are mutually
consistent, but do not support the much larger $S_{2.2}/S_{1.6}$ of
Brown {\it et al}.  We have no explanation for the differences between
our spectrum and that of Brown {\it et al.}

\subsection{Discussion}

We also obtained spectra of 1996 TL$_{66}$ (Fig. 17, taken from Luu and
Jewitt 1998), and 1996 TS$_{66}$ (Fig. 15).  When considered with the
spectrum of 1996 TO$_{66}$ (Brown, Cruikshank and Pendleton 1999) we
see a remarkable diversity of spectral characteristics among the KBOs.
Some (2000 EB$_{173}$, 1996 TL$_{66}$, 1996 TS$_{66}$ and 1993 SC)
appear spectrally bland at the achievable signal-to-noise ratios.
Others (1996 TO$_{66}$, 1999 DE$_9$) show absorptions due to water
ice.  The remarkable 1999 DE$_9$ shows additional evidence for
absorption near $1 \micron$ that may indicate surface olivines and
weaker bands as discussed above.

What do these spectral differences mean?  There is no obvious correlation
with the optical colors (e.g. $B-I$, Table 2) or absolute magnitudes.
Indeed, it is too early to detect patterns in the spectral properties of the
bodies in the outer solar system.  An enlarged sample of high quality
spectra is needed, which will require long integrations on the
brightest KBOs with the largest telescopes.  Parallel measurements of
the albedos using simultaneous optical and thermal measurements
(presumably from the SIRTF spacecraft but perhaps also from
ground-based submillimeter telescopes) are also needed to better
understand the compositions of the KBOs.

The Centaurs and the nuclei of the Jupiter-family comets probably share
a common origin in the Kuiper Belt.  Spectral observations of these
objects should show a diversity parallel to that seen among the KBOs.
Indeed, near infrared spectra of the Centaurs (2060) Chiron and (5145)
Pholus show, respectively, neutral near infrared continuum with
superimposed water ice bands but no other features (Foster {\it et al.}
1999, Luu, Jewitt and Trujillo 2000) and a structured continuum with
evidence for olivine, a low mass hydrocarbon (possibly methanol) as
well as water (Cruikshank {\it et al.} 1999).  Qualitatively, at least,
these two Centaurs are analogues of Kuiper Belt Objects 1996 TO$_{66}$
(Brown, Cruikshank and Pendleton 1999) and 1999 DE$_9$ (Figure 14),
respectively.  Spectra of more Centaurs are urgently needed.  Recent
estimates predict that 20 - 30 Centaurs should have $m_R < 20$ and thus
be within range of spectral investigation with the largest telescopes
(Figure 4 of Sheppard {\it et al.} 2000).  The nuclei of comets present
an even more difficult observational challenge.  They are small and
tend to be surrounded by dust comae when close to the Sun and bright
enough to be studied.  Despite this, obtaining nucleus spectra should
be a high scientific priority for those interested in the origin of the
comets and the nature of the Kuiper Belt.

\section{Summary}

We present new optical photometry and near-infrared spectra of Kuiper
Belt Objects, taken to study the spectral diversity among these bodies
and to search for physically revealing correlations. We find that

\begin{enumerate}
\item The optical (B-V, V-R and R-I) and optical to near-infrared (V-J)
colors of KBOs show a dispersion that is large compared to the
uncertainties of measurement.  The color indices are mutually
correlated, indicating the ubiquitous presence of a single reddening
agent in the $0.45\micron$ to $1.2\micron$ wavelength range.

\item There is no statistically significant ($3\sigma$) evidence for a
bimodal distribution of the B-V vs. V-R colors, either in our sample or
in the observations reported by Tegler and Romanishin (1998, 2000).

\item The colors are not linearly correlated with any orbital or known
physical properties of the KBOs.  Specifically, we find no linear
correlation of the optical color indices with heliocentric distance,
orbital semi-major axis, inclination, eccentricity, circular velocity
difference or absolute magnitude.  However, at the $3\sigma$ confidence
level, we do confirm Tegler and Romanishin's (2000) finding that
objects with perihelia $q >$ 40 AU are systematically redder than
others. The mean colors of the classical and resonant Kuiper Belt
Objects are indistinguishable.

\item  Impact resurfacing (Luu and Jewitt 1996) is probably not the
primary cause of the color dispersion among KBOs, for two reasons.
First, the color dispersion among KBOs is larger than can be easily
produced in Monte Carlo models of the resurfacing process.  Second, the
resurfacing hypothesis predicts that, in a statistical sense,
individual objects should show azimuthal color variations as large as
the color differences between objects.  The measurements show that this
is not the case.

\item The KBOs are also spectrally diverse in the near-infrared, from
featureless to continua marked by distinct solid state absorptions.  In
particular, 1999 DE$_9$ shows remarkable similarity to the spectrum of
a Mauna Kea cinder plus water ice mixture, with a water ice feature
near 2.0 $\micron$ and depression of the continuum at $\lambda \le 1.3
\micron$ that is consistent with (but not uniquely diagnostic of)
olivine absorption.

\item Our $1.4 \micron - 2.4 \micron$ reflection spectrum of 1993 SC
shows no evidence of the existence of absorption features reported by
Brown {\it et al.} (1997).  Furthermore, the spectral slope reported by
Brown et al.  is inconsistent with the spectral slope measured here,
and with independent broadband photometry.

\end{enumerate}

\noindent {\bf Acknowledgements} 

\noindent We thank Wayne Wack, Gary Punawai and John Dvorak for their
observing assistance, and Support Scientists Ken Motohara and Bob
Goodrich for their help with the set-ups.  Chad Trujillo and Scott
Sheppard helped with some of the observations.  Ted Roush and Ed
Cloutis gave helpful comments on our spectrum of 1999 DE$_9$.  John
Davies provided a prompt review.  We thank Jan Kleyna and Jing Li for
discussions about statistics and we gratefully acknowledge support from
NASA's Origins Program.

\newpage

\noindent{\bf Figure Captions}

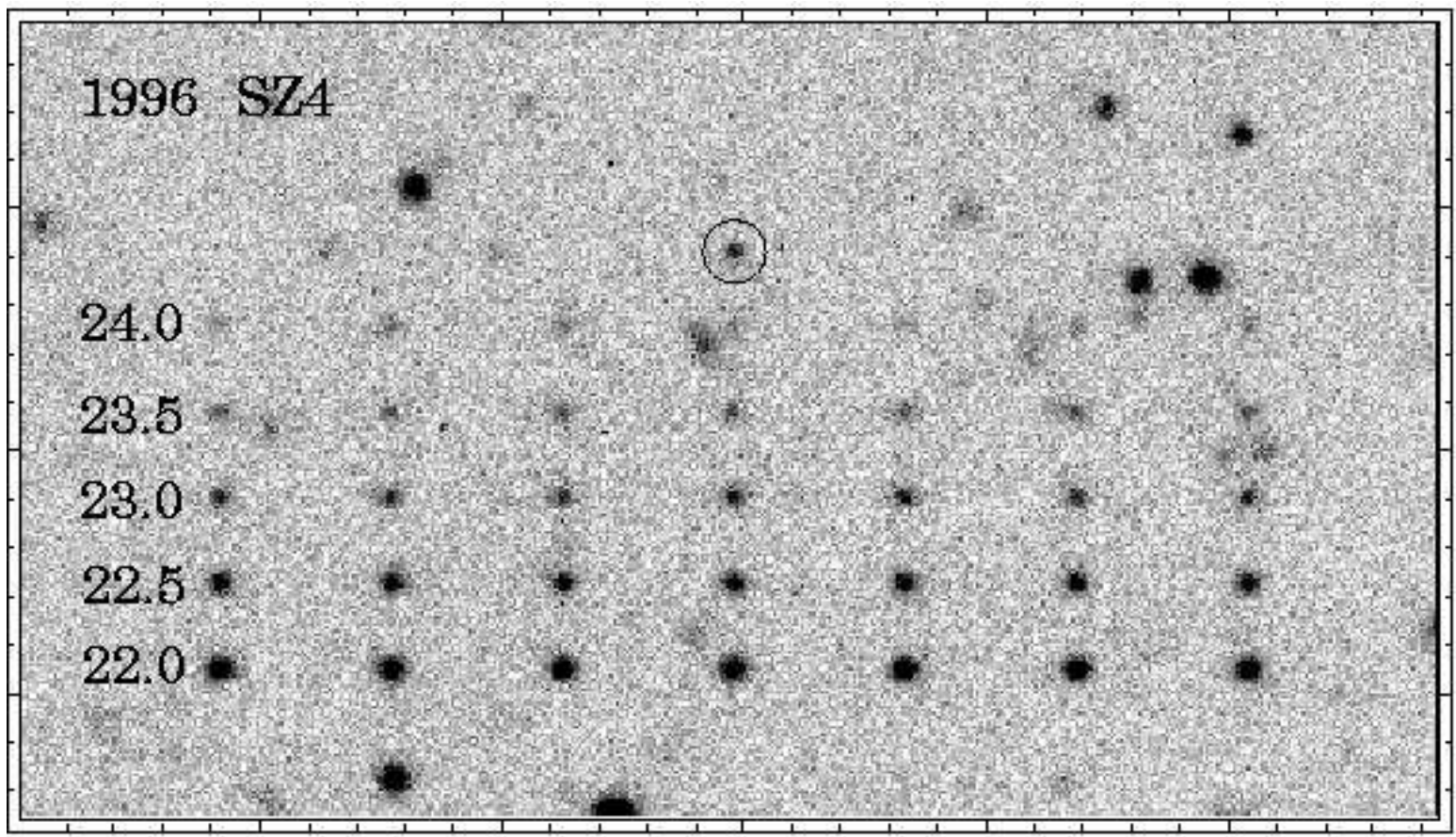
\noindent Figure 1. Keck image of 1996 SZ$_4$ (circled) with a grid of
artificial images added to assess the noise properties of the data.
The magnitudes of the synthesized images vary vertically in the figure,
as marked.  At each magnitude, seven artifical images have been added
to measure the effect of background fluctuations on the extracted
photometry.  This is a 300 second integration through the R filter,
taken 1998 November 14.

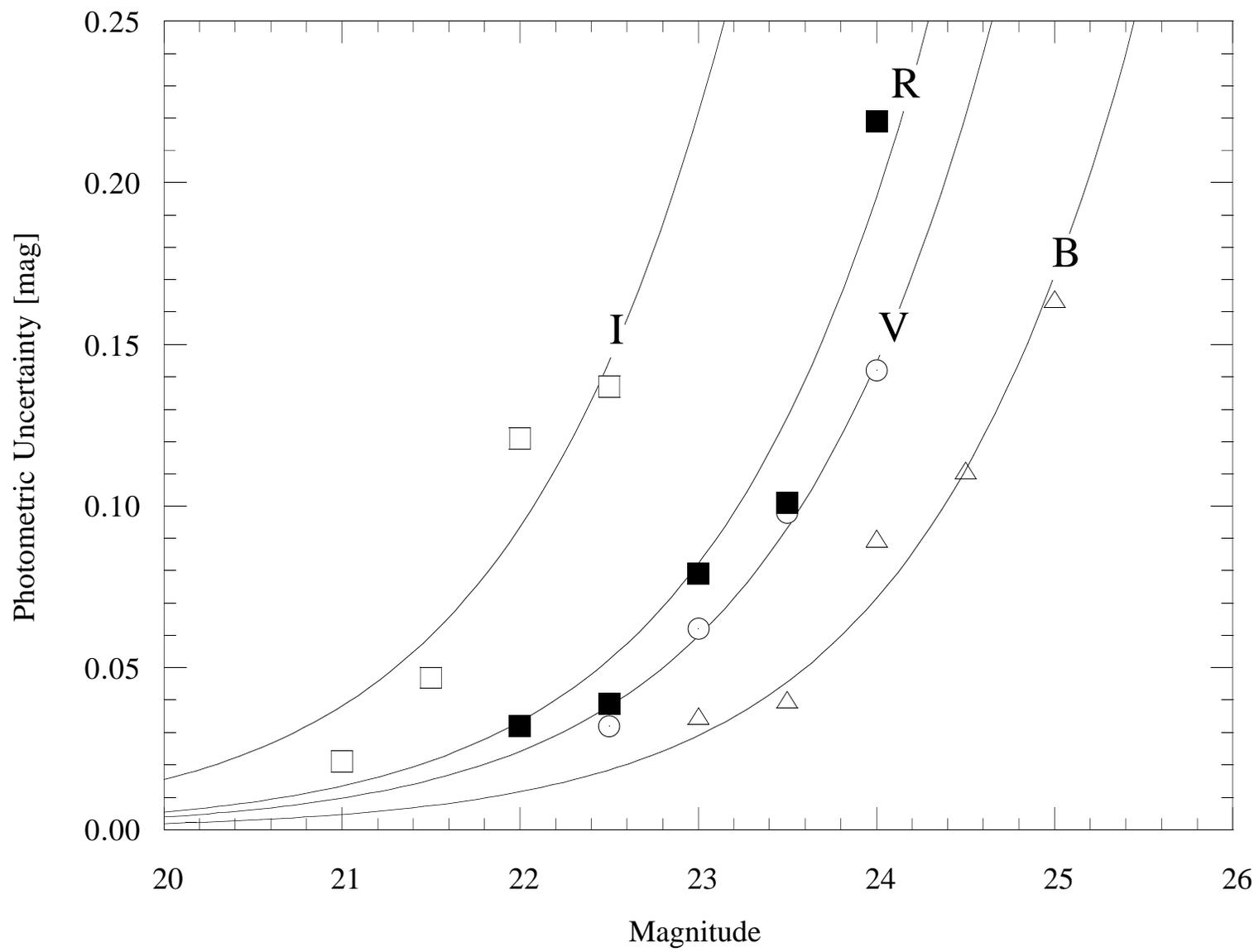
\noindent Figure 2. Sample photometric uncertainties as a function of
apparent magnitude in the BVR and I filters for 300 sec integrations,
measured from simulations like the one in Figure 1. The curves show Eq.
(4) fitted to the data.

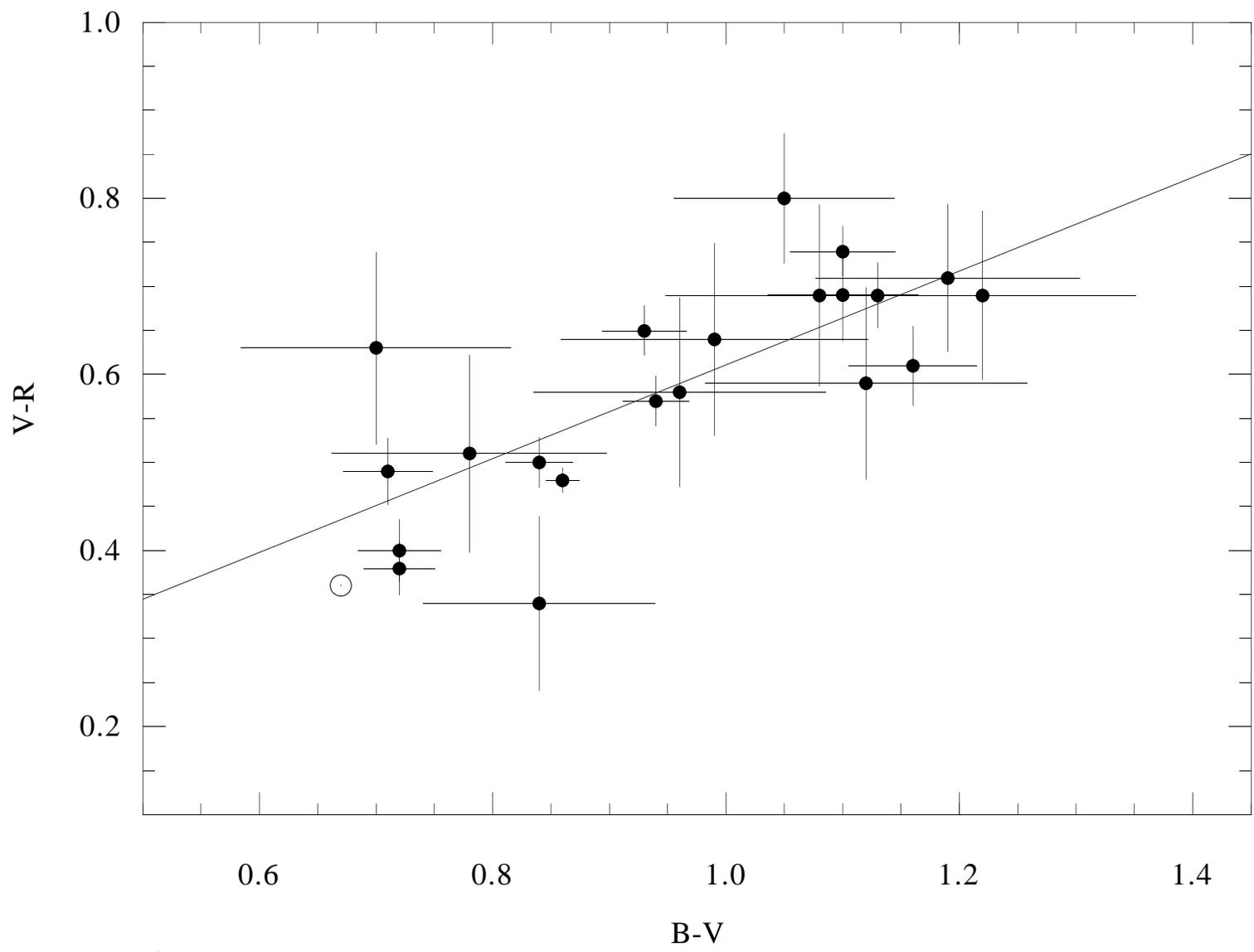
\noindent Figure 3. Plot of B-V vs. V-R for photometry from the present
work having $\sigma_{B-V} < 0.15$ (see Table 2).  The color of the Sun
is marked.  The least squares fit is shown to guide the eye.

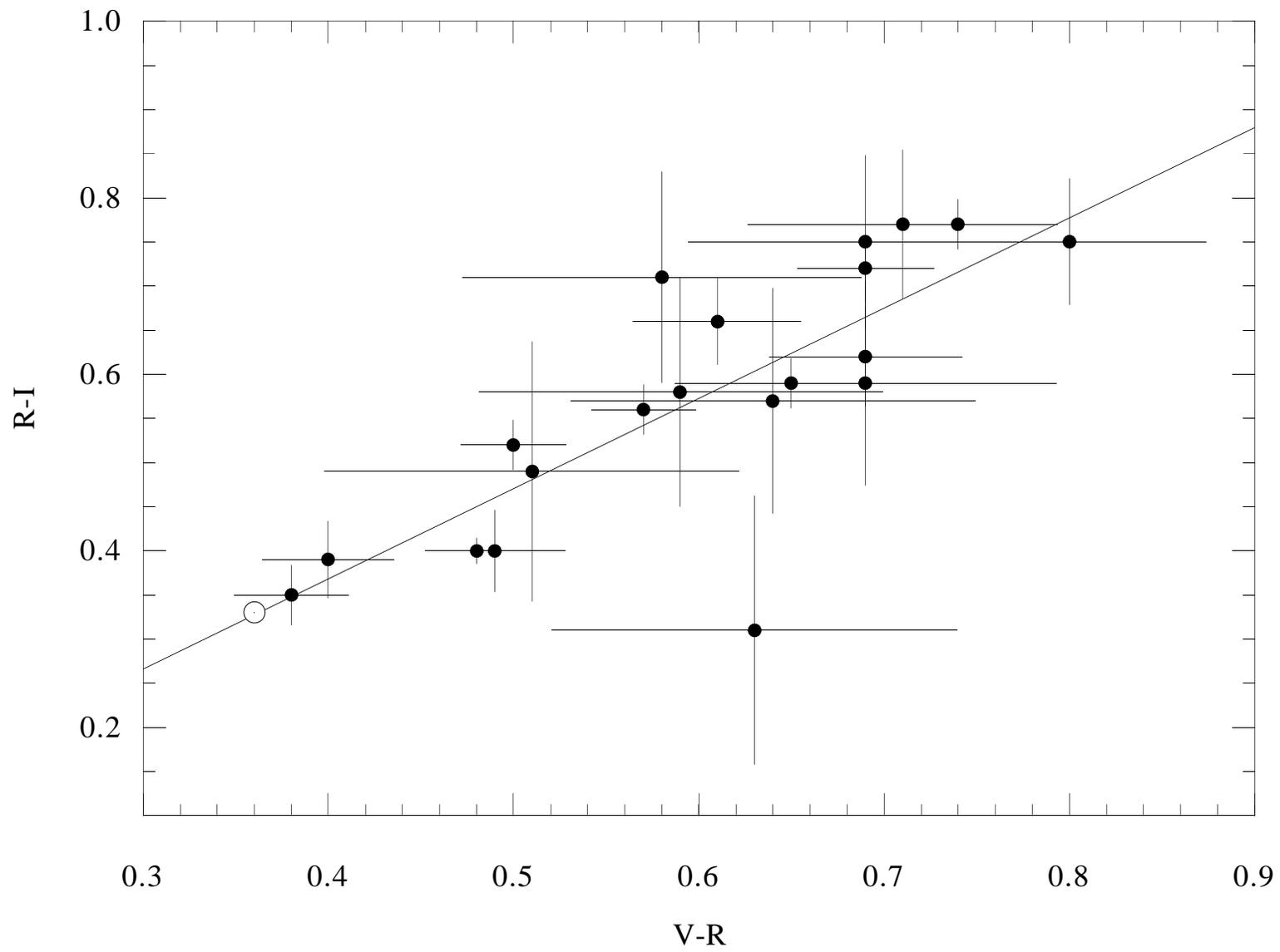
\noindent Figure 4. Plot of V-R vs. R-I for photometry from the present
work having $\sigma_{B-V} < 0.15$ (see Table 2).  The color of the Sun
is marked.

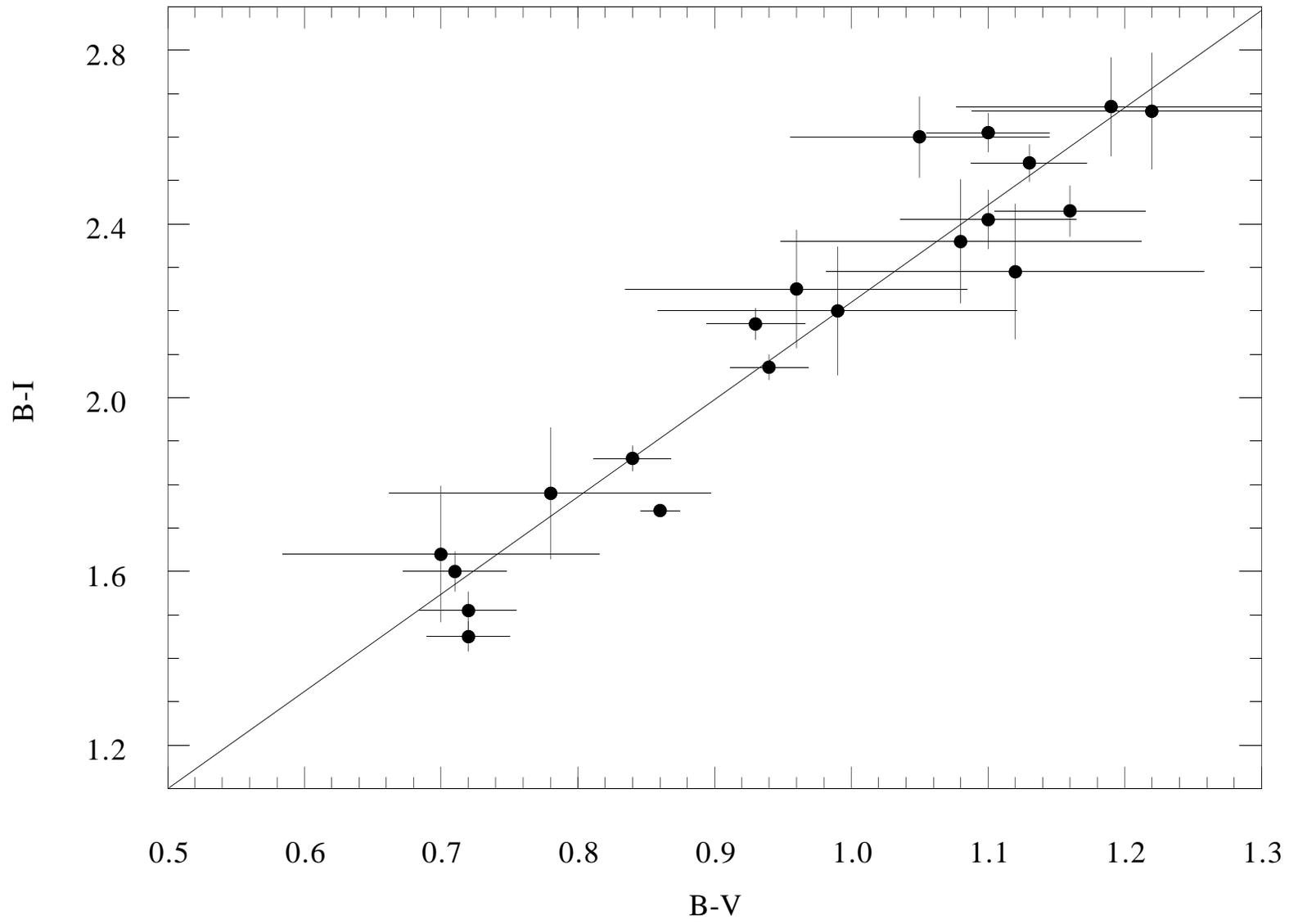
\noindent Figure 5. Plot of B-V vs. B-I for photometry from the present
work having $\sigma_{B-V} < 0.15$ (see Table 2).  The color of the Sun
is marked.

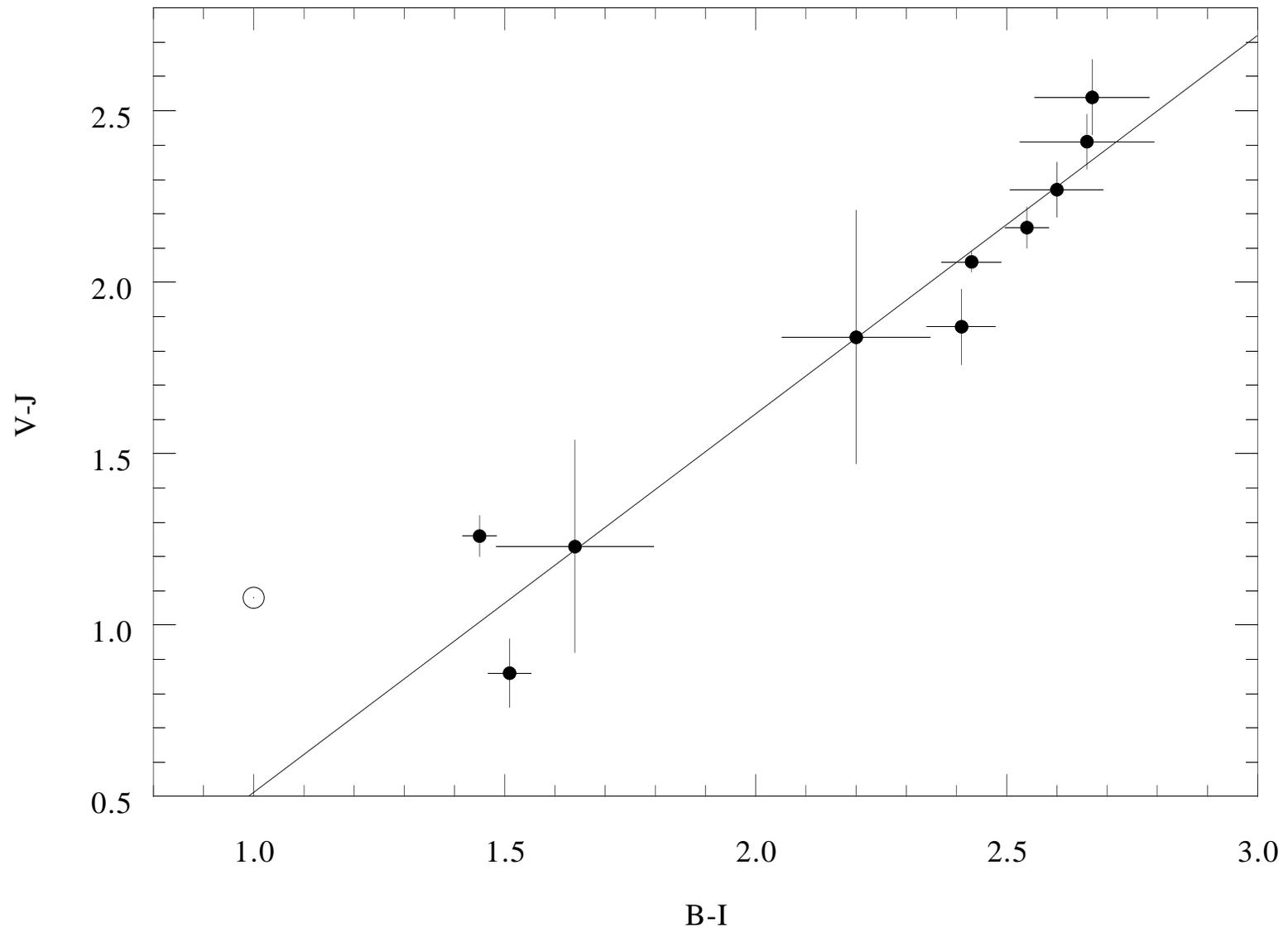
\noindent Figure 6. Plot of B-I vs. V-J for KBOs observed independently
in the near-infrared by Jewitt and Luu (1998) and Davies {\it et al.}
2000.

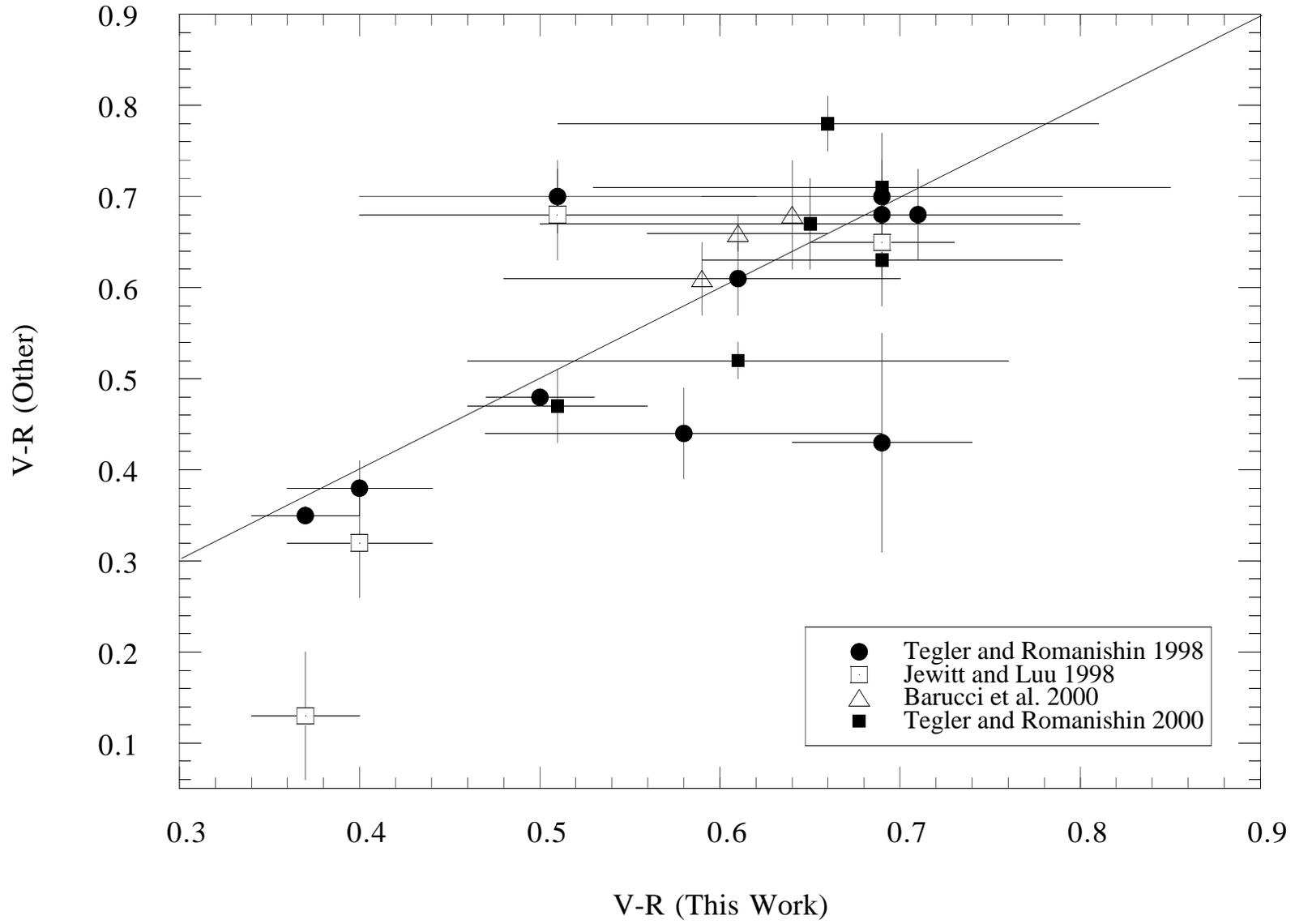
\noindent Figure 7. Comparison of the V-R colors of KBOs measured in
this work and reported independently.  TR98, TR00 are Tegler and
Romanishin 1998 and 2000, respectively, JL98 is Jewitt and Luu 1998
while B00 is Barucci {\it et al.} 2000. The solid line is the track
expected if the measurements are equal.  Error bars have been
suppressed for clarity: they may be read from Table 5.

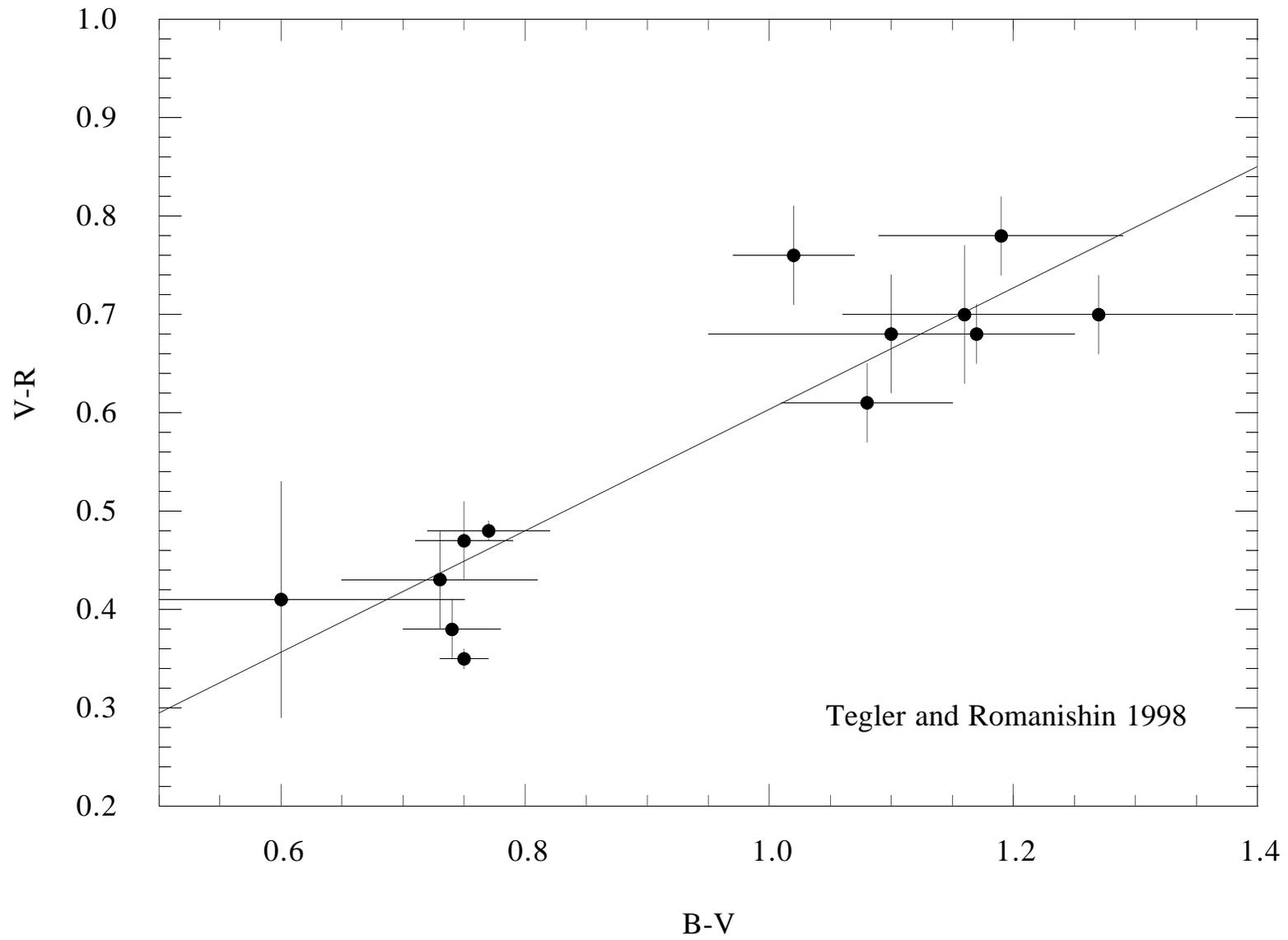
\noindent Figure 8. B-V vs. V-R from Tegler and Romanishin (1998).

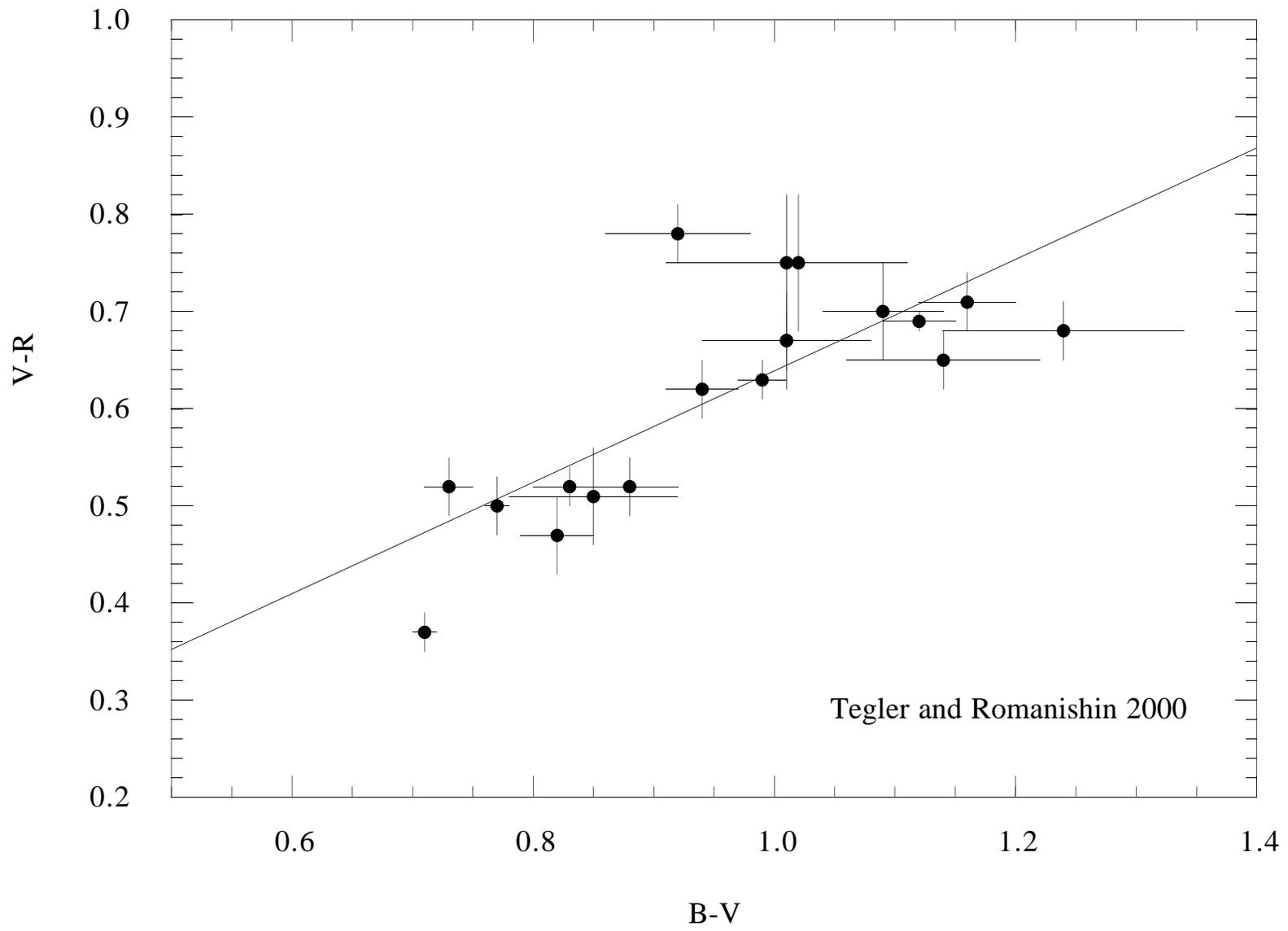
\noindent Figure 9. B-V vs. V-R from Tegler and Romanishin (2000). 

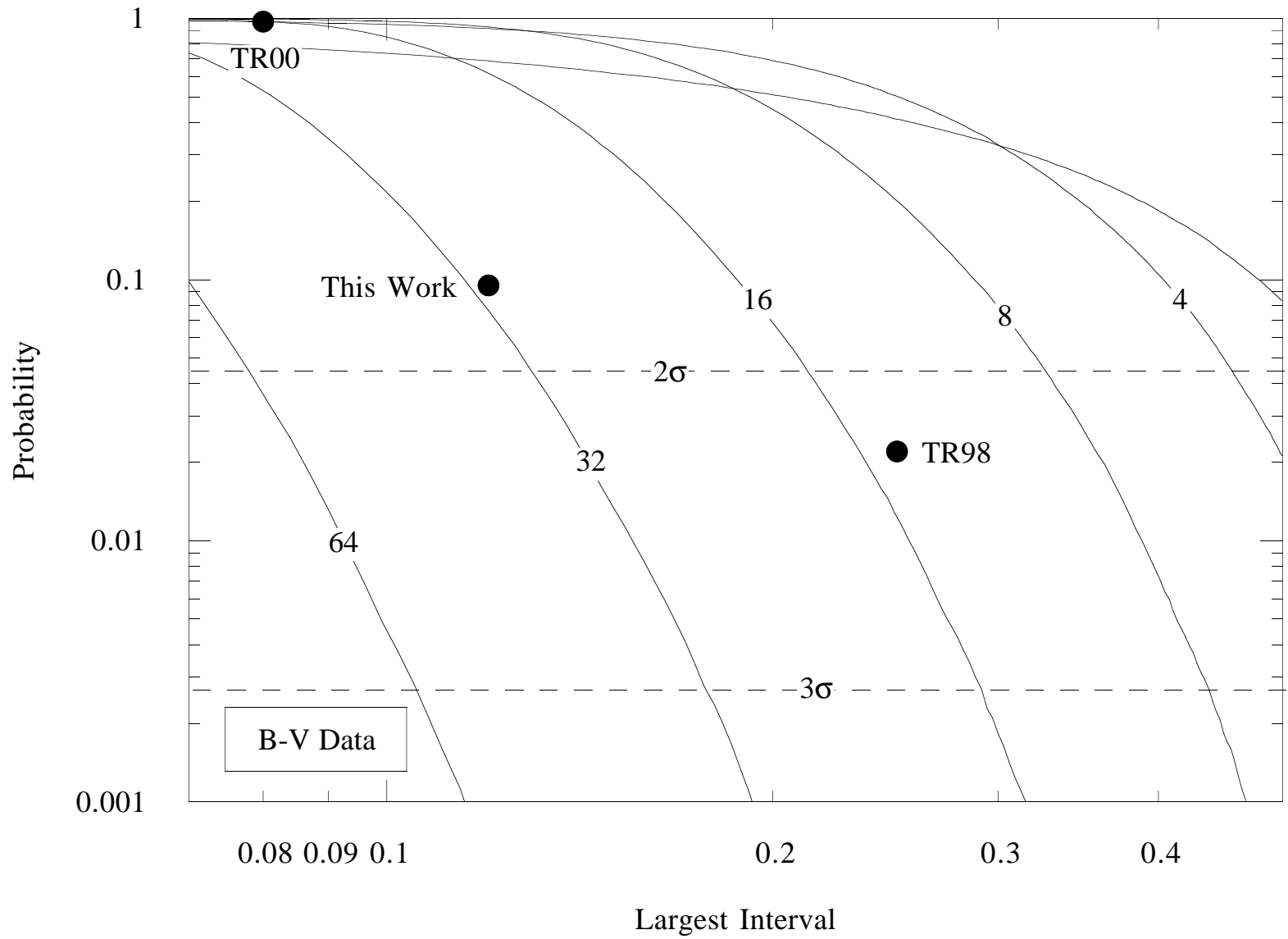
\noindent Figure 10. Probability of obtaining a Largest Interval (LI)
greater than the one observed in a random sampling of a uniform
distribution in $B-V$.  The probabilities were calculated as described
in the text.  Solid lines are marked with the number of data points
included in each simulation.  Dashed lines mark 2$\sigma$ and 3$\sigma$
statistical confidence levels.  Points show the observational samples.

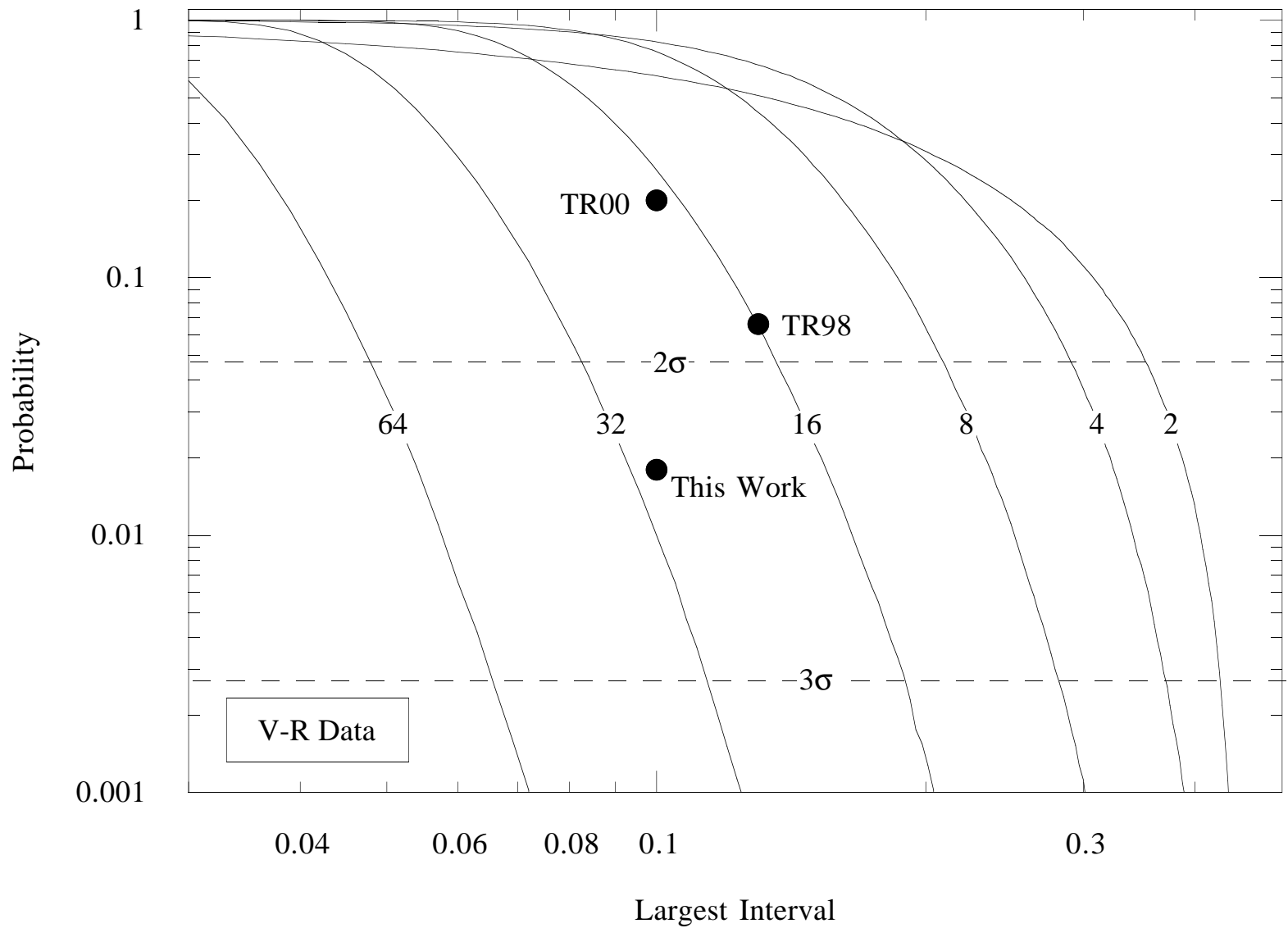
\noindent Figure 11.  Same as Fig. 10 but for the $V-R$ color index.

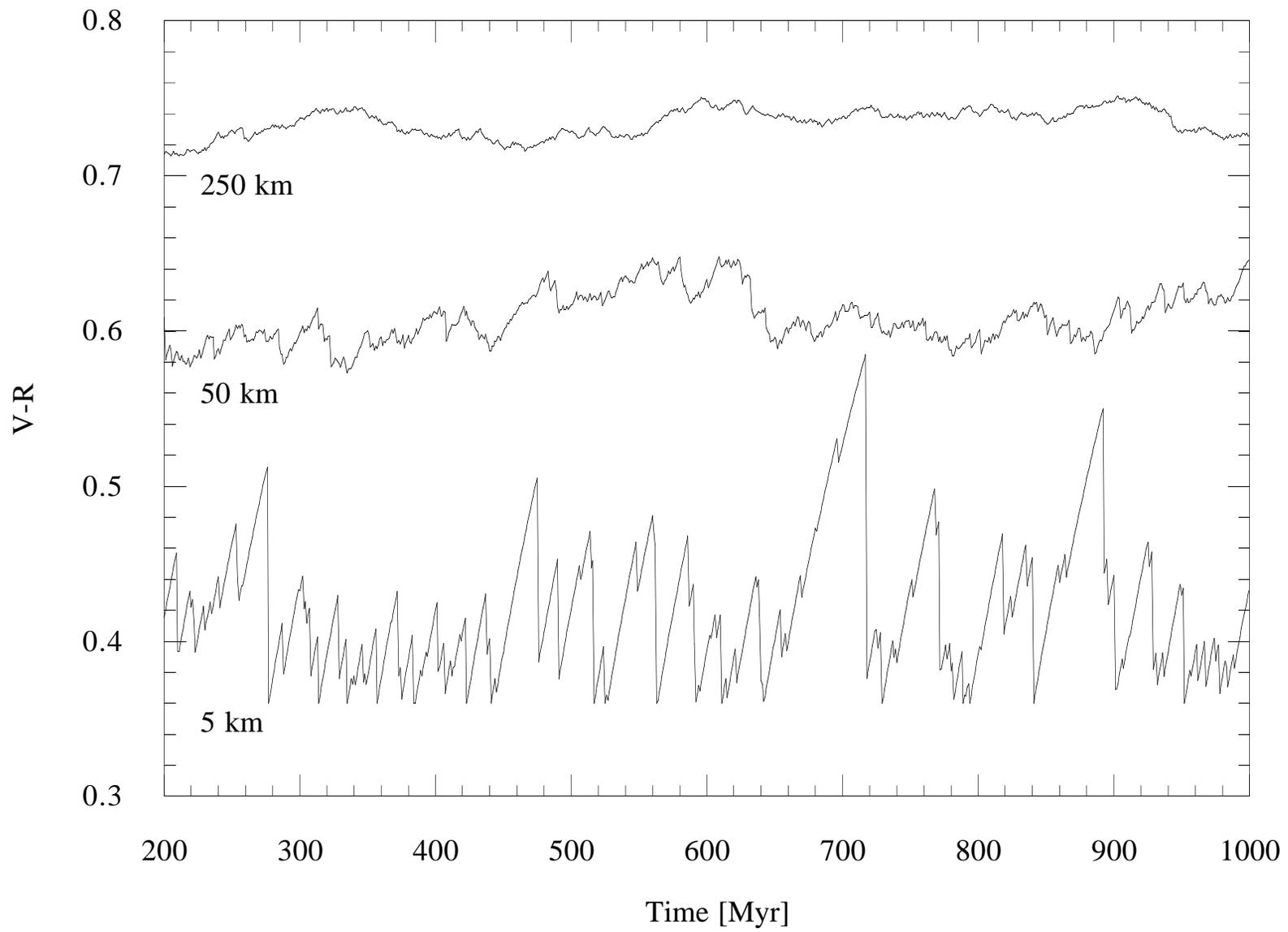
\noindent Figure 12.  Sample resurfacing models computed as described in 
section 3.  The curves show simulated color evolution on three KBOs having 
radii $a_T$ = 5, 50 and 250 km (bottom to top).  A $q = 4$ size distribution
is used, as suggested by the observations, with a minimum projectile radius $a_{min}$ = 0.05 km.  At the right we show the mean colors and the standard
color deviations for each model.

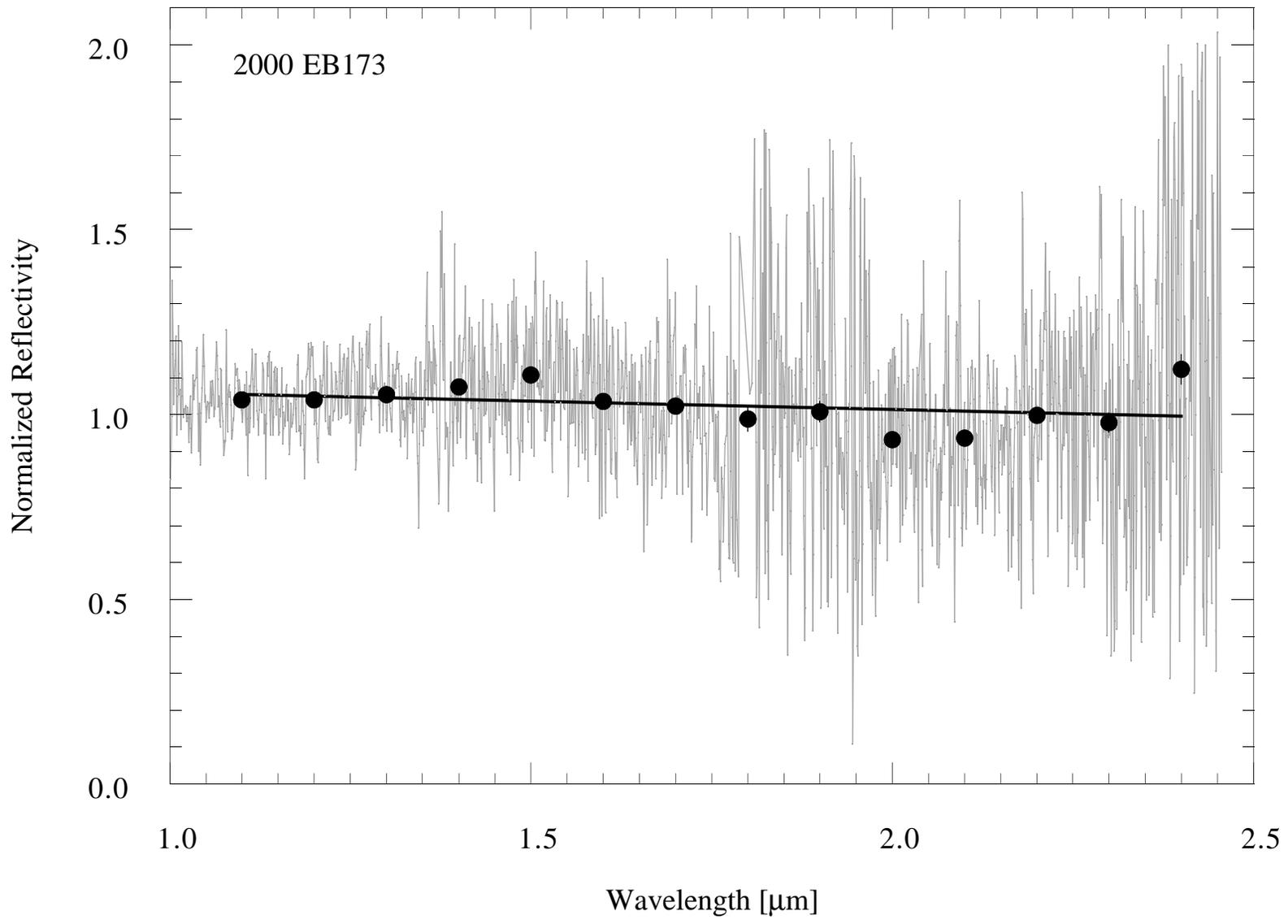
\noindent Figure 13. Reflection spectrum of 2000 EB$_{173}$ taken with
the Subaru telescope. The raw data are plotted in grey.  The black
circles denote the reflectivity binned to $0.1\micron$ resolution (see
also Table 10).  The solid line is a linear, least squares fit to the
data.

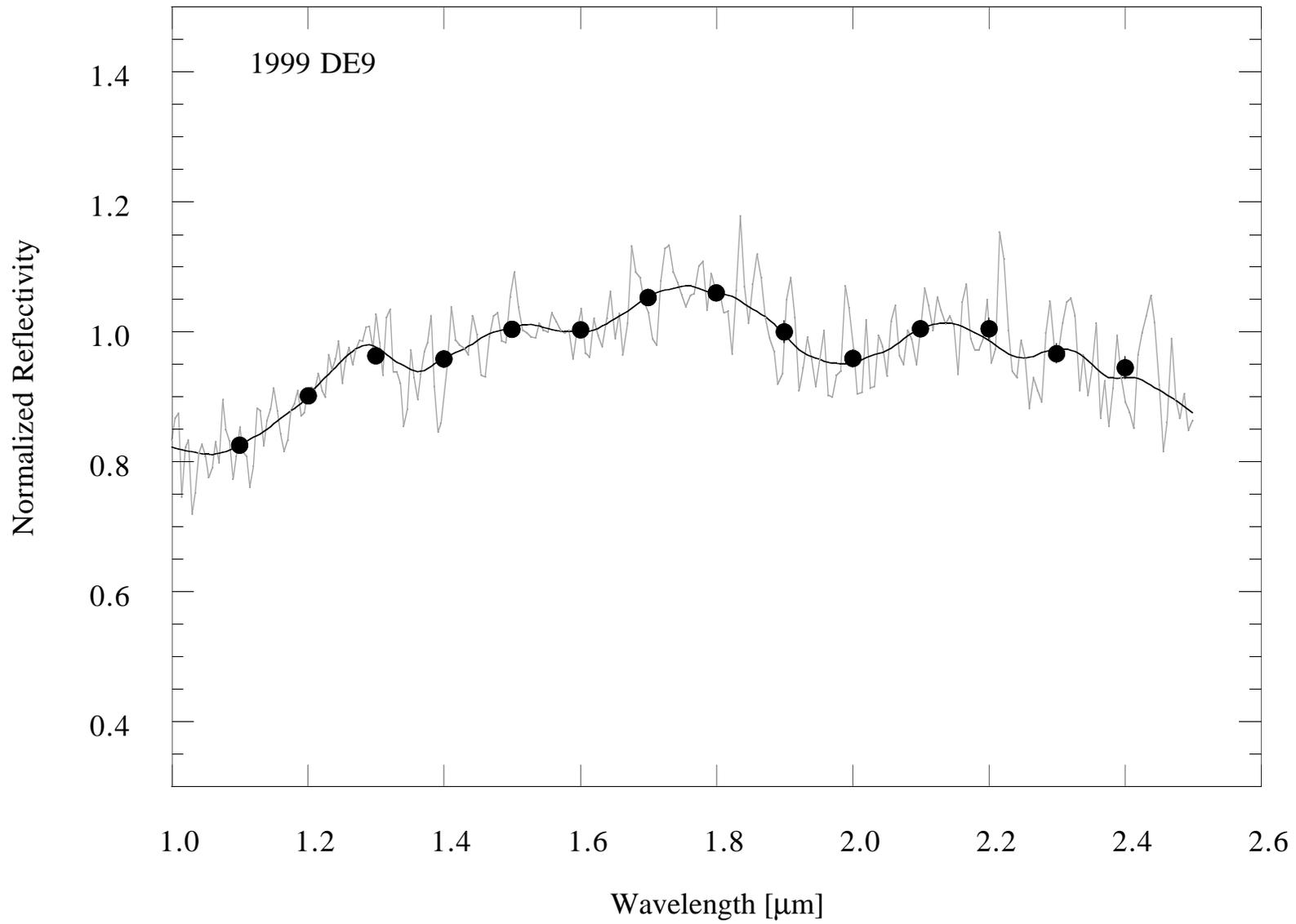
\noindent Figure 14. Reflection spectrum of 1999 DE$_9$ taken with the
Keck telescope.  The grey line shows the raw data.  The black line
through the data is a running mean added to guide the eye.

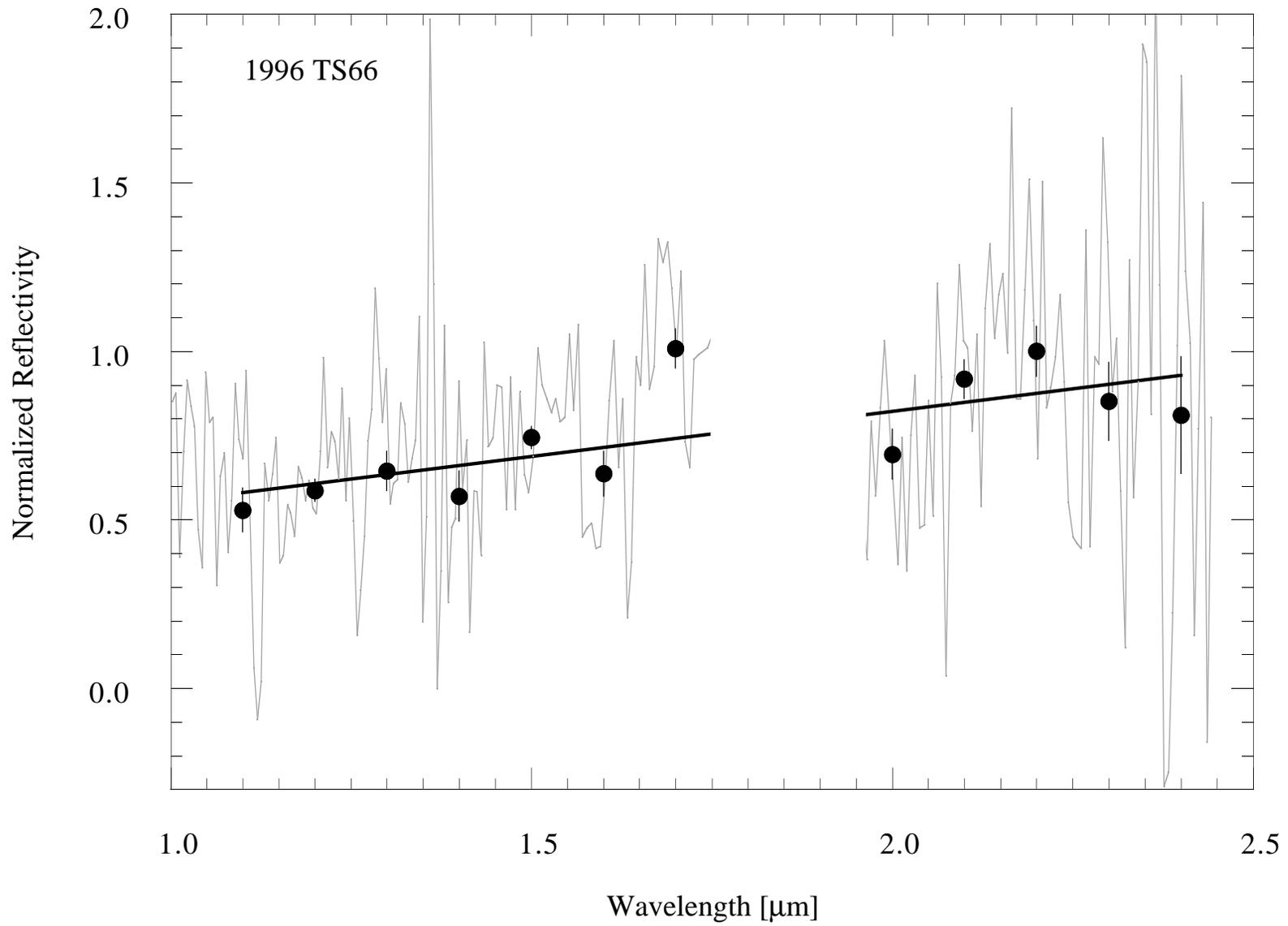
\noindent Figure 15. Reflection spectrum of 1996 TS66 taken with the
Keck telescope. Individual pixels are plotted in grey.  The black
circles denote the reflectivity binned to $0.1\micron$ resolution (see
also Table 10).

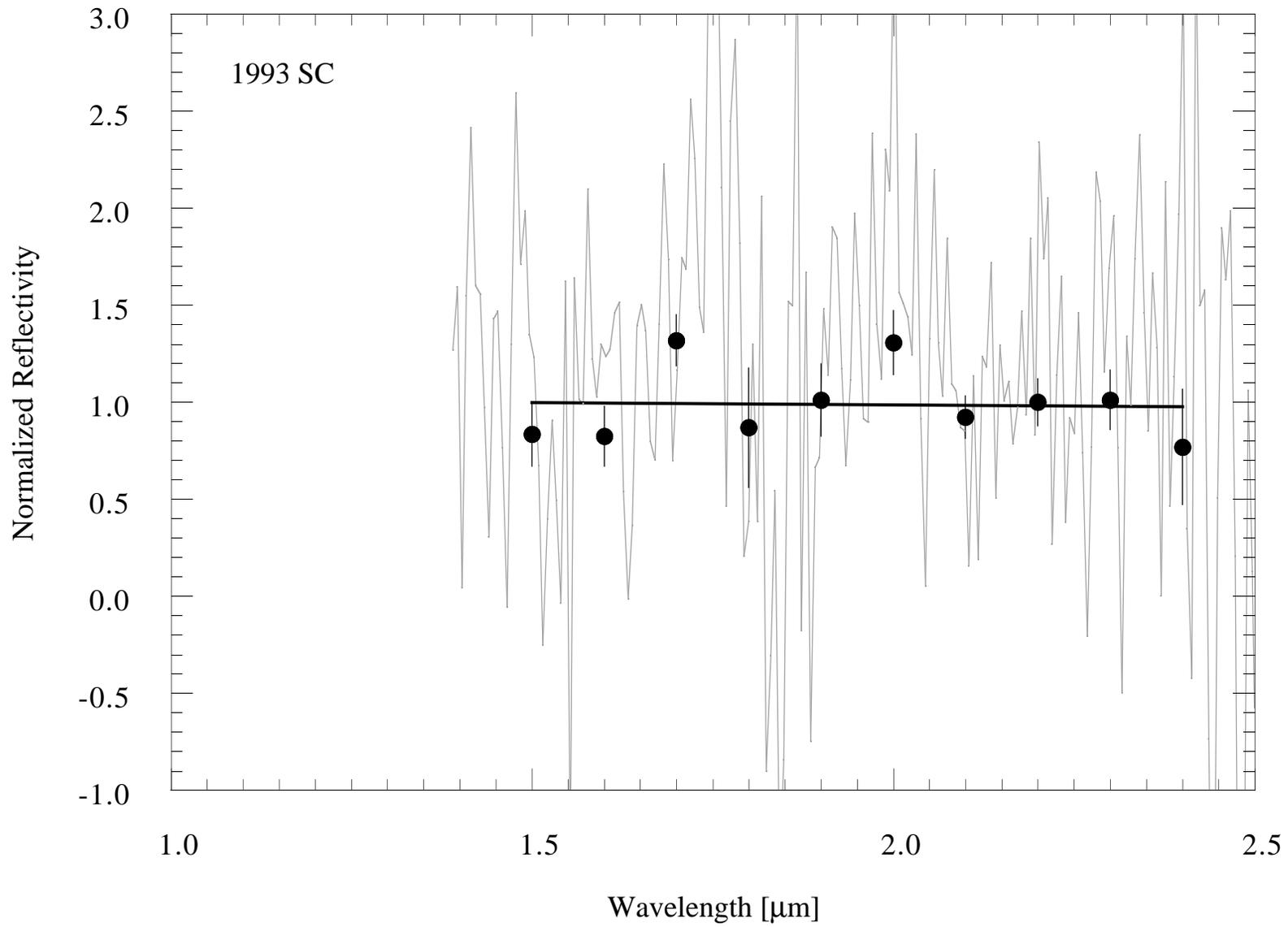
\noindent Figure 16. Same as Figure 12 but for 1993 SC.

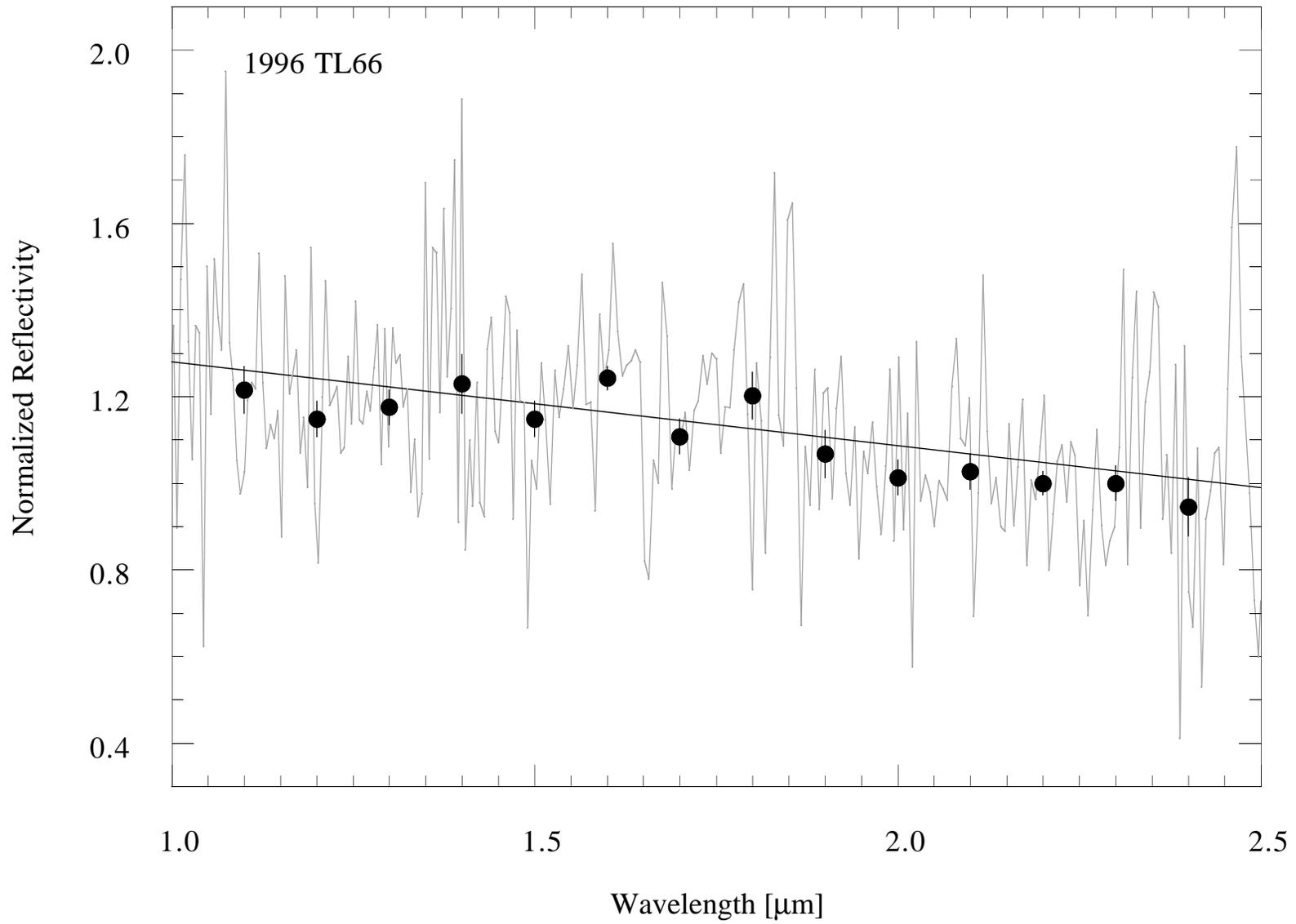
\noindent Figure 17. Same as Figure 12 but for 1996 TL66 (from Jewitt and 
Luu 1998).

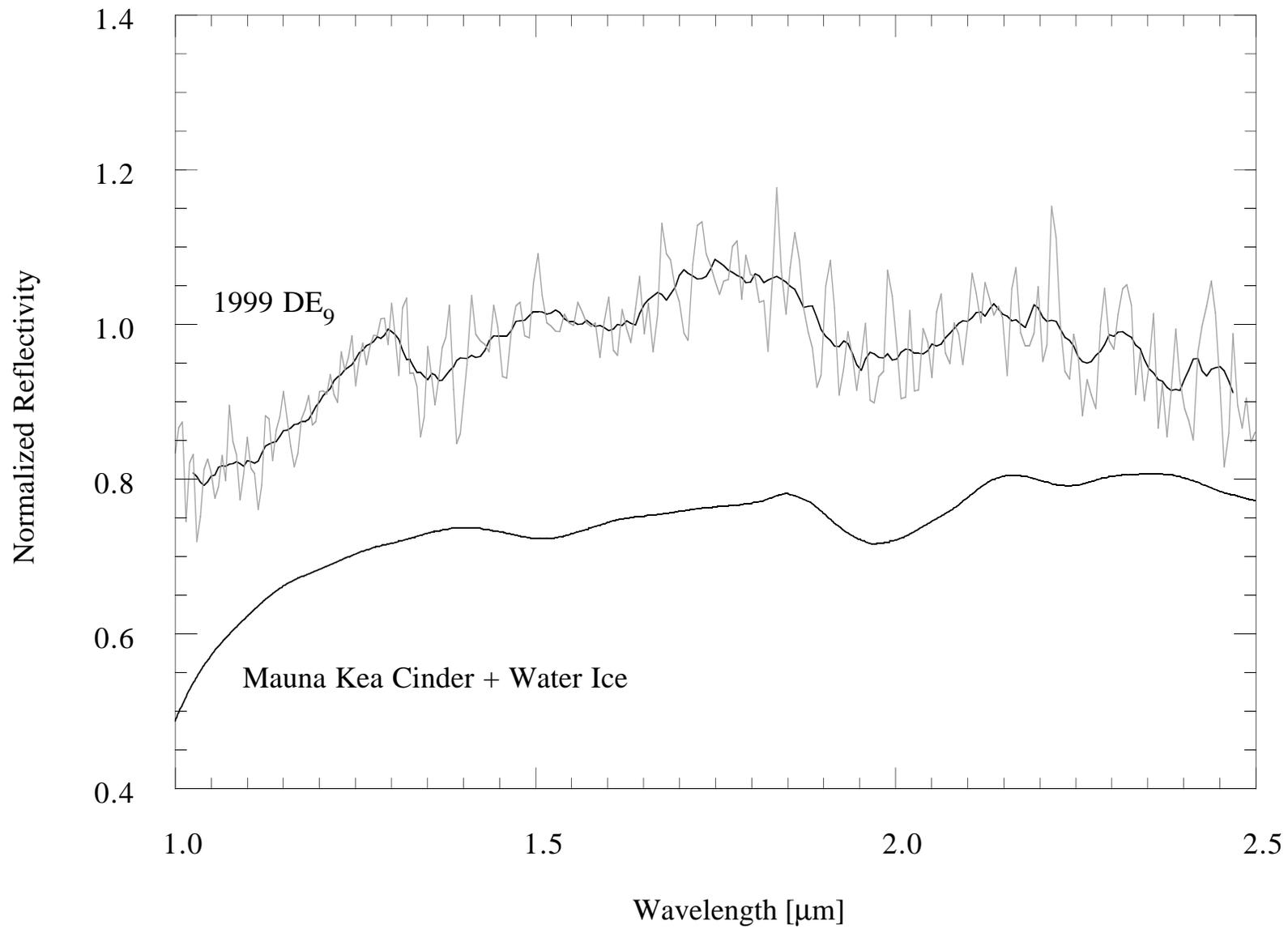
\noindent Figure 18.  Reflection spectrum of 1999 DE$_9$ compared with a
Mauna Kea cinder plus 1\% water ice (by weight) sample from
Clark 1981b.  The 1999 DE$_9$ spectrum is shown unsmoothed (grey line)
and running-box smoothed by 0.05$\micron$ (black line).  The Mauna Kea
spectrum has been vertically offset for clarity.

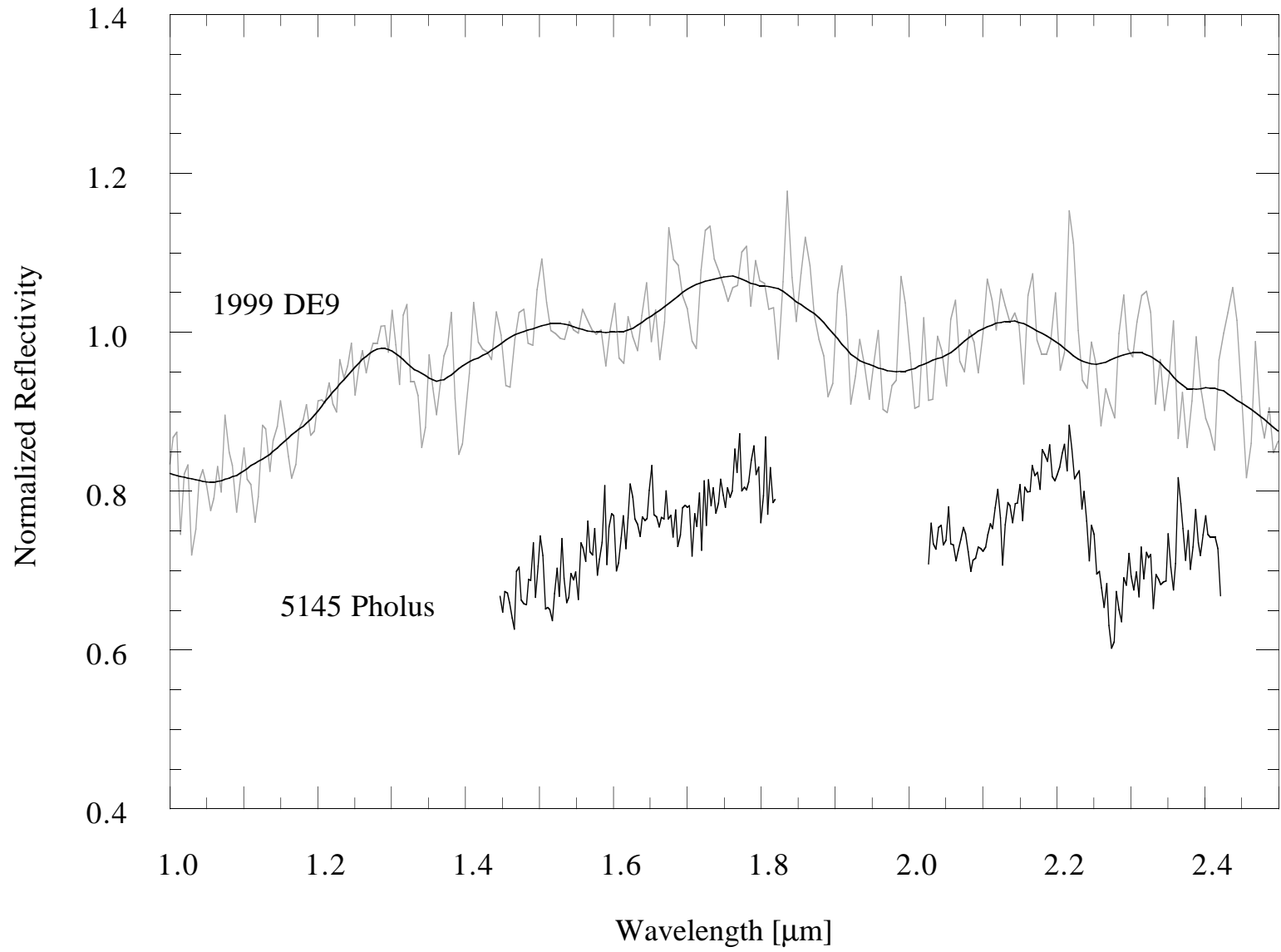
\noindent Figure 19.  Reflection spectrum of 1999 DE$_9$ compared with
the spectrum of Centaur (5145) Pholus (from Luu, Jewitt and Cloutis
1994).  The 1999 DE$_9$ spectrum is shown unsmoothed (grey line) and
running-box smoothed by 0.05$\micron$ (black line). The (5145) Pholus
spectrum has been vertically offset for clarity.

\newpage

\begin{deluxetable}{llllllll}
\footnotesize
\tablecaption{Geometrical Circumstances of the Observations}
\tablewidth{0pc}
\tablehead{
\colhead{Object}&\colhead{Class}&\colhead{Telescope}&
\colhead{UT Date}&\colhead{$R$}&\colhead{$D$}&\colhead{$\alpha$}
\\[.2ex]
\colhead{}&\colhead{}&\colhead{}&\colhead{}&\colhead{[AU]}&\colhead{[AU]}&\colhead{[deg]}
}
\startdata
Pluto&3:2&UH 2.2-m&2000 May 01&30.26&29.41&1.04\\
1992 QB$_1$&CKBO&Keck 10-m&1998 Nov 14&40.91&40.19&0.96\\
1993 SB&3:2&Keck 10-m&1998 Nov 15&31.05&30.3&1.28\\
1993 SC&3:2&Keck 10-m&1998 Nov 14&34.66&33.94&1.14\\
1994 TA&Centaur&Keck 10-m&1998 Nov 14&16.93&15.96&0.74\\
1994 TB&3:2&Keck 10-m&1998 Nov 15&30.08&29.48&1.51\\
1995 DA$_2$&4:3&Keck 10-m&1998 Nov 14&34.06&33.95&1.65\\
1995 WY$_2$&CKBO&Keck 10-m&1998 Nov 15&47.42&46.56&0.58\\
1995 WY$_2$&CKBO&Keck 10-m&1998 Nov 14&47.42&46.57&0.60\\
1996 RQ$_{20}$&CKBO&Keck 10-m&1998 Nov 15&39.46&38.70&0.91\\
1996 RR$_{20}$&3:2&Keck 10-m&1998 Nov 14&43.55&43.17&1.20\\
1996 SZ$_4$&3:2&Keck 10-m&1998 Nov 14&30.24&29.45&1.14\\
1996 TK$_{66}$&CKBO&Keck 10-m&1998 Nov 15&42.84&42.16&0.96\\
1996 TL$_{66}$&SKBO&Keck 10-m&1998 Nov 14&35.09&34.14&0.43\\
1996 TO$_{66}$&CKBO&Keck 10-m&1998 Nov 14&45.86&45.25&0.97\\
1996 TP$_{66}$&3:2&Keck 10-m&1998 Nov 14&26.39&25.49&0.93\\
1996 TQ$_{66}$&3:2&Keck 10-m&1998 Nov 15&34.61&33.70&0.65\\
1996 TS$_{66}$&CKBO&Keck 10-m&1998 Nov 15&38.82&37.87&0.37\\
1997 CQ$_{29}$&CKBO&Keck 10-m&1998 Nov 15&41.28&41.57&1.30\\
1997 CR$_{29}$&2:1?&Keck 10-m&1998 Nov 15&41.85&41.76&1.35\\
1997 CS$_{29}$&CKBO&Keck 10-m&1998 Nov 14&43.6&43.17&1.17\\
1997 CU$_{26}$&Centaur&Keck 10-m&1998 Nov 15&13.55&13.54&4.18\\
1997 CU$_{29}$&CKBO&Keck 10-m&1998 Nov 15&44.75&44.29&1.12\\
1997 QH$_4$&CKBO&Keck 10-m&1998 Nov 15&41.41&40.86&1.14\\
1997 QJ$_4$&3:2&Keck 10-m&1998 Nov 15&34.72&34.16&1.35\\
1998 SN$_{165}$&CKBO&Keck 10-m&1998 Nov 14&38.24&37.75&1.23\\
1999 DE$_9$&SKBO&UH 2.2-m&2000 Apr 28&33.81&33.78&1.54\\
1999 KR$_{16}$&CKBO&UH 2.2-m&2000 Apr 28&38.05&37.07&0.34\\
2000 EB$_{173}$&3:2&UH 2.2-m&2000 Jul 01&29.87&29.73&1.93\\
\enddata
\end{deluxetable}

\begin{deluxetable}{llclccc}
\footnotesize
\tablecaption{Optical Photometry}
\tablewidth{0pc}
\tablehead{
\colhead{Object}&\colhead{Date}&\colhead{$m_R(1,1,0)$}&
\colhead{R}&\colhead{B-V}&\colhead{V-R}&\colhead{R-I}
\\[.2ex]
\colhead{}&\colhead{}&\colhead{[mag]}&\colhead{[mag]}&\colhead{[mag]}&\colhead{[mag]}&\colhead{[mag]}
}
\startdata
Pluto&2000 May 01&-1.37&$13.42 \pm 0.01$&$0.86 \pm 0.01$&$0.48 \pm 0.01$&$0.40
\pm 0.01$\\
1992 QB$_1$&1998 Nov 14&6.98&$23.10 \pm 0.09$&$0.99 \pm 0.18$&$0.66 \pm 0.15$&$0.80 \pm0.15$\\
1993 SB&1998 Nov 15&7.91&$22.83 \pm 0.07$&$0.78 \pm 0.12$&$0.51 \pm 0.11$&$0.49 \pm 0.15$\\
1993 SC&1998 Nov 14&6.73&$22.13 \pm 0.04$&$1.05 \pm 0.10$&$0.80 \pm 0.07$&$0.75 \pm 0.07$\\
1994 TA&1998 Nov 14&11.25&$23.44 \pm 0.12$&$1.36 \pm 0.29$&$0.62 \pm 0.20$&$0.74 \pm 0.21$\\
1994 TB&1998 Nov 15&7.55&$22.35 \pm 0.05$&$1.19 \pm 0.11$&$0.71 \pm 0.08$&$0.77 \pm 0.08$\\
1995 DA$_2$&1998 Nov 14&8.13&$23.51 \pm 0.13$&$1.31 \pm 0.27$&$0.51 \pm 0.20$&$0.63 \pm 0.24$\\
1995 WY$_2$&1998 Nov 15&6.88&$23.62 \pm 0.14$&$1.03 \pm 0.28$&$0.60 \pm 0.23$&$0.51 \pm 0.28$\\
1995 WY$_2$&1998 Nov 14&6.95&$23.69 \pm 0.15$&--&--&--\\
1996 RQ$_{20}$&1998 Nov 15&6.78&$22.74 \pm 0.07$&$0.96 \pm 0.13$&$0.58 \pm 0.11$&$0.71 \pm 0.12$\\
1996 RR$_{20}$&1998 Nov 14&6.72&$23.14 \pm 0.10$&$1.10 \pm 0.21$&$0.69 \pm 0.16$&$0.76 \pm 0.16$\\
1996 SZ$_4$&1998 Nov 14&8.34&$23.13 \pm 0.10$&$0.55 \pm 0.15$&$0.61 \pm 0.15$&$0.62 \pm 0.17$\\
1996 TK$_{66}$&1998 Nov 15&6.30&$22.62 \pm 0.06$&$1.08 \pm 0.13$&$0.69 \pm 0.10$&$0.59 \pm 0.12$\\
1996 TL$_{66}$&1998 Nov 14&5.04&$20.45 \pm 0.02$&$0.72 \pm 0.03$&$0.38 \pm 0.03$&$0.35 \pm 0.03$\\
1996 TO$_{66}$&1998 Nov 14&4.49&$21.11 \pm 0.03$&$0.72 \pm 0.03$&$0.40 \pm 0.04$&$0.39 \pm 0.04$\\
1996 TP$_{66}$&1998 Nov 14&6.85&$21.03 \pm 0.02$&$1.13 \pm 0.04$&$0.69 \pm 0.04$&$0.72 \pm 0.04$\\
1996 TQ$_{66}$&1998 Nov 15&7.17&$22.53 \pm 0.06$&$1.22 \pm 0.13$&$0.69 \pm 0.10$&$0.75 \pm 0.10$\\
1996 TS$_{66}$&1998 Nov 15&5.86&$21.71 \pm 0.03$&$1.10 \pm 0.06$&$0.69 \pm 0.05$&$0.62 \pm 0.06$\\
1997 CQ$_{29}$&1998 Nov 15&6.50&$22.72 \pm 0.07$&$0.99 \pm 0.13$&$0.64 \pm 0.11$&$0.57 \pm 0.13$\\
1997 CR$_{29}$&1998 Nov 15&7.04&$23.31 \pm 0.11$&$0.67 \pm 0.20$&$0.69 \pm 0.18$&$0.51 \pm 0.22$\\
1997 CS$_{29}$&1998 Nov 14&5.11&$21.53 \pm 0.03$&$1.16 \pm 0.06$&$0.61 \pm 0.05$&$0.66 \pm 0.05$\\
1997 CU$_{26}$&1998 Nov 15&6.39&$17.87 \pm 0.02$&$0.84 \pm 0.03$&$0.50 \pm 0.03$&$0.52 \pm 0.03$\\
1997 CU$_{29}$&1998 Nov 15&6.22&$22.75 \pm 0.07$&$1.12 \pm 0.14$&$0.59 \pm 0.11$&$0.58 \pm 0.13$\\
1997 QH$_4$&1998 Nov 15&6.88&$23.07 \pm 0.09$&$1.05 \pm 0.18$&$0.65 \pm 0.15$&$0.64 \pm 0.16$\\
1997 QJ$_4$&1998 Nov 15&7.31&$22.73 \pm 0.07$&$0.70 \pm 0.12$&$0.63 \pm 0.11$&$0.31 \pm 0.15$\\
1998 SN$_{165}$&1998 Nov 14&5.37&$21.22 \pm 0.03$&$0.71 \pm 0.04$&$0.40 \pm 0.04$&$0.40 \pm 0.05$\\
1999 DE$_9$&2000 Apr 28&4.68&$20.04 \pm 0.02$&$0.94 \pm 0.03$&$0.57 \pm 0.03$&$0.56 \pm 0.03$\\
1999 KR$_{16}$&2000 Apr 28&5.49&$21.25 \pm 0.02$&$1.10 \pm 0.05$&$0.74 \pm 0.03$&$0.77 \pm 0.03$\\
2000 EB$_{173}$&2000 Jul 01&4.61&$19.43 \pm 0.02$&$0.93 \pm 0.04$&$0.65 \pm 0.03$&$0.59 \pm 0.03$\\
\enddata
\end{deluxetable}

\begin{deluxetable}{cccccc}
\footnotesize
\tablecaption{Correlations}
\tablewidth{0pc}
\tablehead{
\colhead{Quantity\tablenotemark{1}}&\colhead{Quantity\tablenotemark{1}}
&\colhead{$N$\tablenotemark{2}}&
\colhead{$r_{corr}$\tablenotemark{3}}
&\colhead{$P(r \ge r_{corr})$\tablenotemark{4}}
}
\startdata
B-I&V-J&10&0.97&$<0.001$\\
B-V&R-I&28&0.70&$<0.001$\\
V-R&R-I&28&0.71&$<0.001$\\
B-V&V-R&27&0.44&0.02\\
B-I&$m_R(1,1,0)$&28&0.38&0.05\\
B-I&$a$&28&0.21&$>0.1$\\
B-I&$e$&28&0.30&$>0.1$\\
B-I&$i$&28&-0.08&$>0.1$\\
B-I&$m_R$&28&0.35&$0.05$\\
B-I&$q$&28&0.18&$>0.1$\\
B-I&$R_h$&28&0.02&$>0.1$\\
B-I&$\Delta V$&28&0.19&$>0.1$\\
\enddata
\noindent \tablenotetext{1}{$m_R(1,1,0) =$ red magnitude reduced to
$R = \Delta = 1 AU$ and to phase angle $\alpha = 0\deg$. $a =$
semi-major axis, $q =$ perihelion distance, $R_h = $ heliocentric distance,
$\Delta V$ = velocity relative to uninclined, circular Keplerian orbit}
\noindent \tablenotetext{2}{Number of measurements in the sample}
\noindent \tablenotetext{3}{Linear correlation coefficient}
\noindent \tablenotetext{4}{Probability that a larger correlation
coefficient could be obtained by chance from $N$ measurements of uncorrelated
data.  $P < 0.003$ indicates a correlation with $> 3\sigma$ significance.}
\end{deluxetable}

\begin{deluxetable}{ccccccc}
\footnotesize
\tablecaption{Color Means}
\tablewidth{0pc}
\tablehead{
\colhead{Sample\tablenotemark{1}}&\colhead{$N$\tablenotemark{2}}
&\colhead{$B-V$\tablenotemark{3}}&
\colhead{$V-R$\tablenotemark{3}}
&\colhead{$R-I$\tablenotemark{3}}
&\colhead{$B-I$\tablenotemark{3}}
}
\startdata
Classical&12&$1.00\pm0.04$&$0.61\pm0.03$&$0.60\pm0.04$&$2.22\pm0.10$\\
All Resonant&12&$0.96\pm0.07$&$0.64\pm0.03$&$0.61\pm0.04$&$2.20\pm0.12$\\
3:2 Only&10&$0.95\pm0.07$&$0.65\pm0.03$&$0.62\pm0.05$&$2.21\pm0.14$\\
All&28&$0.98\pm0.04$&$0.61\pm0.02$&$0.60\pm0.03$&$2.18\pm0.07$\\
\enddata
\noindent \tablenotetext{1}{Dynamical sub-sample.}
\noindent \tablenotetext{2}{Number of measurements in the sub-sample}
\noindent \tablenotetext{3}{Mean color and standard deviation on the mean.}
\end{deluxetable}

\begin{deluxetable}{llllll}
\small
\footnotesize
\tablecaption{Comparison of Optical Measurements}
\tablewidth{0pc}
\tabletypesize{\footnotesize}
\tablehead{
\colhead{Object}&\colhead{Source}&\colhead{$m_R(1,1,0)$}&
\colhead{B-V}&\colhead{V-R}&\colhead{R-I}\\[.2ex]
\colhead{}&\colhead{}&\colhead{[mag]}&\colhead{[mag]}&\colhead{[mag]}&\colhead{[mag]}
}
\startdata
1992 QB$_1$&(This work)&$6.98 \pm 0.09$&$0.99 \pm 0.18$&$0.66 \pm 0.15$&$0.80 \pm 0.15$\\
&(TR00)&$6.92 \pm 0.05$&$0.92 \pm 0.06$&$0.78 \pm 0.03$&--\\
1993 SB&(This work)&$7.91 \pm 0.07$&$0.78 \pm 0.12$&$0.51 \pm 0.11$&$0.49 \pm 0.15$\\
&(TR00)&$7.80 \pm 0.04$&$0.82 \pm 0.03$&$0.47 \pm 0.04$&--\\
1993 SC&(This work)&$6.73 \pm 0.04$&$1.05 \pm 0.09$&$0.80 \pm 0.07$&$0.75 \pm 0.07$\\
&(LJ96)&$6.56 \pm 0.05$&$0.92 \pm 0.11$&$0.57 \pm 0.09$&$0.86 \pm 0.10$\\
&DMG97&$6.49 \pm 0.10$&--&$0.54 \pm 0.14$&$0.43 \pm 0.14$\\
&(JL98)&$6.71 \pm 0.02$&$0.94 \pm 0.06$&$0.68 \pm 0.05$&$0.68 \pm 0.04$\\
&(TR98)&--&$1.27 \pm 0.11$&$0.70 \pm 0.04$&--\\
1994 TB&(This work)&$7.55 \pm 0.05$&$1.19 \pm 0.11$&$0.71 \pm 0.08$&$0.77 \pm 0.08$\\
&(TR98)&--&$1.10 \pm 0.15$&$0.68 \pm 0.06$&--\\
1996 CU$_{26}$&(This work)&$6.39 \pm 0.02$&$0.84 \pm 0.03$&$0.50 \pm 0.03$&$0.52 \pm 0.03$\\
&(TR98)&--&$0.77 \pm 0.05$&$0.48 \pm 0.01$&--\\
&(M99)&$6.18 \pm 0.05$&--&$0.46 \pm 0.02$&$0.56 \pm 0.03$\\
1996 RQ$_{20}$&(This work)&$6.78 \pm 0.07$&$0.96 \pm 0.13$&$0.58 \pm 0.11$&$0.71 \pm 0.12$\\
&(TR98)&--&--&$0.44 \pm 0.05$&--\\
1996 RR$_{20}$&(This work)&$6.72 \pm 0.10$&$1.10 \pm 0.21$&$0.69 \pm 0.16$&$0.76 \pm 0.16$\\
&(TR00)&$6.51 \pm 0.03$&$1.16 \pm 0.04$&$0.71 \pm 0.03$&--\\
1996 SZ$_4$&(This work)&$8.34 \pm 0.10$&$0.55 \pm 0.15$&$0.61 \pm 0.15$&$0.62 \pm 0.17$\\
&(TR00)&$8.04 \pm 0.04$&$0.83 \pm 0.03$&$0.52 \pm 0.02$&--\\
1996 TK$_{66}$&(This work)&$6.30 \pm 0.06$&$1.08 \pm 0.13$&$0.69 \pm 0.10$&$0.59 \pm 0.12$\\
&(TR00)&$6.21 \pm 0.04$&$0.99 \pm 0.02$&$0.63 \pm 0.02$&--\\
1996 TL$_{66}$&(This work)&$5.04 \pm 0.02$&$0.72 \pm 0.03$&$0.37 \pm 0.03$&$0.36 \pm 0.04$\\
&(JL98)&$5.32 \pm 0.04$ &$0.58 \pm 0.05$&$0.13 \pm 0.07$&$0.54 \pm 0.04$\\
&(TR98)&--&$0.75 \pm 0.02$&$0.35 \pm 0.01$&--\\
1996 TO$_{66}$&(This work)&$4.49 \pm 0.03$&$0.72 \pm 0.04$&$0.40 \pm 0.04$&$0.39 \pm 0.04$\\
&(JL98)&$4.52 \pm 0.05$&$0.59 \pm 0.06$&$0.32 \pm 0.06$&$0.36 \pm 0.07$\\
&(TR98)&--&$0.74 \pm 0.04$&$0.38 \pm 0.03$&--\\
1996 TP$_{66}$&(This work)&$6.85 \pm 0.02$&$1.13 \pm 0.04$&$0.69 \pm 0.04$&$0.72 \pm 0.04$\\
&(JL98)&$6.97 \pm 0.04$&$0.80 \pm 0.08$&$0.65 \pm 0.07$&$0.69 \pm 0.04$\\
&(TR98)&--&$1.17 \pm 0.05$&$0.68 \pm 0.03$&--\\
1996 TQ$_{66}$&(This work)&$7.17 \pm 0.06$&$1.22 \pm 0.13$&$0.69 \pm 0.10$&$0.75 \pm 0.10$\\
&(TR98)&--&$1.16 \pm 0.10$&$0.70 \pm 0.07$&--\\
1996 TS$_{66}$&(This work)&$5.86 \pm 0.03$&$1.10 \pm 0.06$&$0.69 \pm 0.05$&$0.62 \pm 0.06$\\
&(JL98)&$6.11 \pm 0.08$&$0.93 \pm 0.09$&$0.43 \pm 0.12$&$0.67 \pm 0.12$\\
1997 CQ$_{29}$&(This work)&$6.60 \pm 0.07$&$0.99 \pm 0.13$&$0.64 \pm 0.11$&$0.57 \pm 0.13$\\
&(B00)&$6.70 \pm 0.05$&$0.99 \pm 0.12$&$0.68 \pm 0.06$&$0.62 \pm 0.09$\\
1997 CS$_{29}$&(This work)&$5.11 \pm 0.03$&$1.16 \pm 0.05$&$0.61 \pm 0.05$&$0.66 \pm 0.05$\\
&(TR98)&--&$1.08 \pm 0.07$&$0.61 \pm 0.04$&--\\
&(B00)&$4.88 \pm 0.02$&$1.05 \pm 0.06$&$0.66 \pm 0.02$&$0.53 \pm 0.04$\\
1997 CU$_{29}$&(This work)&$6.22 \pm 0.07$&$1.12 \pm 0.14$&$0.59 \pm 0.11$&$0.58 \pm 0.13$\\
&(B00)&$6.16 \pm 0.03$&$1.32 \pm 0.12$&$0.61 \pm 0.04$&$0.74 \pm 0.06$\\
1997 QH$_4$&(This work)&$6.88 \pm 0.09$&$1.05 \pm 0.18$&$0.65 \pm 0.15$&$0.64 \pm 0.16$\\
&(TR00)&$6.88 \pm 0.06$&$1.01 \pm 0.07$&$0.67 \pm 0.05$&--\\
2000 EB$_{173}$&(This work)&$4.61 \pm 0.02$&$0.93 \pm 0.04$&$0.65 \pm 0.03$&$0.59 \pm 0.03$\\
&(F01)&$4.72 \pm 0.06$&$0.99 \pm 0.14$&$0.60 \pm 0.10$&$0.38 \pm 0.09$\\
\enddata
\tablecomments{References are as follows: B00 = Barucci {\it et al.} (2000); 
DMG97 = Davies, McBride and Green (1997); F01 = Ferrin {\it et al.} (2001); 
LJ96 = Luu and Jewitt (1996); M99 = McBride {\it et al.} (1999); TR98 = 
Tegler and Romanishin (1998); TR00 = Tegler and Romanishin (2000)}
\end{deluxetable}

\begin{deluxetable}{cccccccc}
\footnotesize
\tablecaption{Results of Bin Test}
\tablewidth{0pc}
\tablehead{
\colhead{Sample}&\colhead{$n$\tablenotemark{a}}&\colhead{$C_{min}$\tablenotemark{b}}&\colhead{$C_{max}$\tablenotemark{c}}&\colhead{$C \pm \sigma_C$\tablenotemark{d}}&\colhead{$m$\tablenotemark{e}}&\colhead{$P(n,m)$\tablenotemark{f}}&\colhead{$S$\tablenotemark{g}}\\[.2ex]
}
\startdata
TR98&13&0.73&1.45&$1.11\pm0.08$&1&0.033&0.967\\
TR00&19&0.80&1.41&$1.14\pm0.04$&8&0.133&0.867\\
TR98+TR00&32&0.73&1.45&$1.13\pm0.05$&8&0.127&0.873\\
This Work&28&0.81&1.49&$1.16\pm0.04$&10&0.150&0.850\\
\enddata
\noindent \tablenotetext{a}{Number of data points in the sample}
\noindent \tablenotetext{b}{Minimum value of color index}
\noindent \tablenotetext{c}{Maximum value of color index}
\noindent \tablenotetext{d}{Mean and standard deviation on the mean}
\noindent \tablenotetext{e}{Number of objects in central bin}
\noindent \tablenotetext{f}{Probability that $m$ objects would be found in the central color bin from a sample of $n$ objects drawn from a uniform distribution, from Equation 7}
\noindent \tablenotetext{g}{Statistical significance $S = 1 - P(n,m)$.  Note that $S = 0.997$ corresponds to $3\sigma$ confidence.}
\end{deluxetable}

\begin{deluxetable}{ccccc}
\footnotesize
\tablecaption{Results of Dip Test}
\tablewidth{0pc}
\tablehead{
\colhead{Sample}&\colhead{Color}&\colhead{$n$\tablenotemark{a}}&\colhead{Dip statistic}&\colhead{S\tablenotemark{b}}\\[.2ex]
}
\startdata
TR98&B-V&13&0.13736&0.983\\
TR98&V-R&16&0.13542&0.993\\
TR00&B-V&19&0.05263&0.040\\
TR00&V-R&21&0.09921&0.933\\
This Work&B-V&28&0.06766&0.584\\
This Work&V-R&28&0.05555&0.239\\
This Work&R-I&28&0.06790&0.597\\
\enddata
\noindent \tablenotetext{a}{Number of data points}
\noindent \tablenotetext{b}{Statistical significance $S = 0.9970$ corresponds to the nominal $3\sigma$ criterion
for a statistically significant bimodality}
\end{deluxetable}

\begin{deluxetable}{ccccc}
\footnotesize
\tablecaption{Results of Interval Distribution Test}
\tablewidth{0pc}
\tablehead{
\colhead{Sample}&\colhead{Color}&\colhead{$n$\tablenotemark{a}}&\colhead{$LI_{o}$\tablenotemark{b}}&\colhead{$P(LI > LI_{o})$\tablenotemark{c}}\\[.2ex]
}
\startdata
TR98&$B-V$&13&0.25&0.012\\
TR98&$V-R$&16&0.13&0.062\\
TR00&$B-V$&19&0.08&0.974\\
TR00&$V-R$&21&0.10&0.245\\
This Work&$B-V$&28&0.12&0.924\\
This Work&$V-R$&28&0.10&0.990\\
This Work&$R-I$&28&0.09&0.878\\
\enddata
\noindent \tablenotetext{a}{Number of data points}
\noindent \tablenotetext{b}{Largest Interval in the sample}
\noindent \tablenotetext{c}{Probability that a Largest Interval greater than 
the one observed would be produced by chance from data selected at random from
a uniform distribution. $P = 0.003$ corresponds to the nominal $3\sigma$ criterion for 
statistical significance.}
\end{deluxetable}

\begin{deluxetable}{ccccccccc}
\footnotesize
\tablecaption{Observational Parameters of Spectra}
\tablewidth{0pc}
\tablehead{
\colhead{UT Date}&\colhead{Instrument}&\colhead{Object}&
\colhead{$\lambda$}&\colhead{Slit Width}&\colhead{$\lambda/\Delta \lambda$}
&\colhead{Seeing}&\colhead{$\tau_{CSO}$}&\colhead{Int\tablenotemark{a}}\\[.2ex]
\colhead{}&\colhead{}&\colhead{}&\colhead{[$\micron$]}&\colhead{[arcsec]}&\colhead{}&\colhead{[arcsec]}&\colhead{}&\colhead{[sec]}
}
\startdata
{\rm KECK I} &&&&&&&&\\
\\
1998-11-13&NIRC&1993 SC&HK&0.68&$\sim$100&0.6&0.06&6000\\
1998-11-13&NIRC&1996 TS$_{66}$&JH&0.68&$\sim$100&0.6-0.7&0.22&3600\\
1998-11-13&NIRC&1996 TS$_{66}$&HK&0.68&$\sim$100&0.6-0.7&0.22&3600\\
1999-04-3,4&NIRC&1999 DE$_9$&JH&0.68&$\sim$100&0.6&0.10-0.04&3000\\
1999-04-3,4&NIRC&1999 DE$_9$&HK&0.68&$\sim$100&0.6&0.10-0.04&6000\\
\\
{\rm SUBARU} &&&&&&&&\\
\\
2000-06-18&CISCO&2000 EB$_{173}$&JH&1.0&$\sim$1000&0.4&0.07&1440\\
2000-06-18&CISCO&2000 EB$_{173}$&HK&1.0&$\sim$1000&0.4&0.07&2880\\
\enddata
\noindent \tablenotetext{a}{Accumulated integration time}
\end{deluxetable}

\begin{deluxetable}{ccccccc}
\footnotesize
\tablecaption{Infrared Reflectivities}
\tablewidth{0pc}
\tablehead{
\colhead{$\lambda$\tablenotemark{a}}&\colhead{1999 DE$_9$}&\colhead{2000 EB$_{173}$}&
\colhead{1996 TL$_{66}$}&\colhead{1996 TS$_{66}$}&\colhead{1993 SC}
\\[.2ex]
}
\startdata
1.1&$0.820 \pm 0.009$&$1.040 \pm 0.007$&$1.216 \pm 0.054$&$0.529 \pm 0.066$&--
\\
1.2&$0.902 \pm 0.010$&$1.040 \pm 0.008$&$1.149 \pm 0.041$&$0.587 \pm 0.033$&--&\\
1.3&$0.963 \pm 0.011$&$1.053 \pm 0.009$&$1.176 \pm 0.041$&$0.645 \pm 0.058$&--&\\
1.4&$0.958 \pm 0.013$&$1.075 \pm 0.015$&$1.230 \pm 0.068$&$0.570 \pm 0.074$&--&\\
1.5&$1.004 \pm 0.010$&$1.107 \pm 0.013$&$1.149 \pm 0.041$&$0.744 \pm 0.033$&$0.835 \pm 0.165$&\\
1.6&$1.003 \pm 0.007$&$1.036 \pm 0.014$&$1.243 \pm 0.027$&$0.636 \pm 0.066$&$0.824 \pm 0.154$&\\
1.7&$1.053 \pm 0.013$&$1.023 \pm 0.015$&$1.108 \pm 0.041$&$1.008 \pm 0.058$&$1.319 \pm 0.132$&\\
1.8&$1.060 \pm 0.012$&$0.988 \pm 0.034$&$1.203 \pm 0.054$&$0.587 \pm 0.107$&$0.868 \pm 0.308$&\\
1.9&$1.000 \pm 0.017$&$1.007 \pm 0.028$&$1.068 \pm 0.054$&$0.496 \pm 0.107$&$1.011 \pm 0.187$&\\
2.0&$0.959 \pm 0.013$&$0.932 \pm 0.021$&$1.014 \pm 0.041$&$0.694 \pm 0.074$&$1.308 \pm 0.165$&\\
2.1&$1.005 \pm 0.009$&$0.936 \pm 0.015$&$1.027 \pm 0.041$&$0.917 \pm 0.058$&$0.923 \pm 0.110$&\\
2.2&$1.000 \pm 0.016$&$1.000 \pm 0.018$&$1.000 \pm 0.027$&$1.000 \pm 0.074$&$1.000 \pm 0.121$&\\
2.3&$0.967 \pm 0.015$&$0.978 \pm 0.026$&$1.000 \pm 0.041$&$0.851 \pm 0.116$&$1.011 \pm 0.154$&\\
2.4&$0.945 \pm 0.017$&$1.122 \pm 0.040$&$0.946 \pm 0.068$&$0.810 \pm 0.174$&$0.769 \pm 0.297$&\\
\enddata
\noindent \tablenotetext{a}{Central wavelength of the bin ($\micron$).}
\end{deluxetable}

\end{document}